\documentclass[DIV12,10pt]{scrartcl}
\usepackage[utf8]{inputenc}
\usepackage[english]{babel}

\usepackage{amsmath,blkarray}

\usepackage{amsthm,amssymb,amsmath}
\usepackage[noblocks]{authblk}

\usepackage{multirow}
\usepackage{multicol}
\usepackage{graphicx}

\usepackage{comment}
\usepackage{caption}
\usepackage{color}
\usepackage{setspace}
\newcommand{\argmax}{\operatornamewithlimits{argmax}}

\setkomafont{title}{\normalfont \LARGE \bfseries}
\setkomafont{section}{\normalfont \Large \bfseries \boldmath}
\setkomafont{subsection}{\normalfont \large \bfseries}
\setkomafont{subsubsection}{\normalfont \normalsize \bfseries}
\setkomafont{descriptionlabel}{\itshape}
\setkomafont{paragraph}{\normalfont \bfseries \boldmath}

\newtheorem{theorem}{Theorem}

\newtheorem{proposition}[theorem]{Proposition}

\usepackage{fancyhdr}
\fancypagestyle{myplain}
{
  \fancyhf{}

  \fancyfoot[C]{\thepage}
}
\title{A Branch--and--Cut Algorithm to Design LDPC Codes without Small Cycles in Communication Systems}

\author[1]{Banu Kabakulak\thanks{Corresponding author. Tel.: +90 2123596771; fax: +90 2122651800. \\ E-mail addresses: banu.kabakulak@boun.edu.tr (B. Kabakulak), caner.taskin@boun.edu.tr (Z. C. Ta\c{s}k\i n), ali.pusane@boun.edu.tr (A. E. Pusane).}}
\author[1]{Z. Caner Ta\c{s}k\i n}
\author[2]{Ali Emre Pusane}

\affil[1]{Department of Industrial Engineering, Bo\u{g}azi\c{c}i University, \.{I}stanbul, Turkey}
\affil[2]{Department of Electrical and Electronics Engineering, Bo\u{g}azi\c{c}i University, \.{I}stanbul, Turkey}

\date{\vspace*{-2em}}

\begin{document}

\maketitle

\thispagestyle{empty}

\begin{abstract}

\vspace{-2mm}

In a digital communication system, information is sent from one place to another over a noisy communication channel using binary symbols (bits). Original information is encoded by adding redundant bits, which are then used by low--density parity--check (LDPC) codes to detect and correct errors that may have been introduced during transmission. Error correction capability of an LDPC code is severely degraded due to harmful structures such as small cycles in its bipartite graph representation known as Tanner graph (TG). 
We introduce an integer programming formulation to generate a TG for a given smallest cycle length. 
We propose a branch-and-cut algorithm for its solution and investigate structural properties of the problem to derive valid inequalities and variable fixing rules. We introduce a heuristic to obtain feasible solutions of the problem. Our computational experiments show that our algorithm can generate LDPC codes without small cycles in acceptable amount of time for practically relevant code lengths. 


\end{abstract}

\newpage
\section{Introduction and Literature Review} \label{Introduction and Literature Review}

Telecommunication is the transmission of messages from a transmitter to a receiver over a potentially unreliable communication environment. In a digital communication system, binary code symbols (\emph{bits}) represent the messages.
In parallel to the rapid developments in technology, digital communication systems find several application areas: messaging via digital cellular phones, fiber optic internet, TV broadcasting or agricultural monitoring through digital satellites, and receiving high quality images of Jupiter under NASA's Juno mission \cite{MJ} are some examples of digital communication.

In practice, numerous transmitter--receiver pairs share the same communication environment such as air or space. Hence, radio waves, electrical signals, and  light waves over fiber optic channels accumulate some amount of noise on the medium. The noise in the environment can cause transmission errors or failures. Channel coding is the term used for the collection of techniques that are employed in digital communications to ensure that a transmission is recovered with minimal or no errors. These techniques encode the original information by adding redundant bits. When the receiver receives information, the decoder estimates the original information by detecting and correcting errors in the received vector with the help of redundant bits. 

Among the codes that are used in the decoding process at receiver, low--density parity--check (LDPC) code family has received attention thanks to its high error detection and correction capabilities. LDPC codes were first proposed by Gallager in 1962 and today they are used in wireless network standard (IEEE 802.11n), WiMax (IEEE 802.16e), and digital video broadcasting standard (DVB-S2) \cite{G62}. They have sparse parity--check matrices, i.e., $\mathbf{H}$ matrix, and can alternatively be represented by bipartite graphs known as Tanner graphs (TG) \cite{T81}. A TG (or LDPC code) is said to be \emph{(J, K)--regular} if all nodes at one side of the bipartite graph have degree $J$ and all other nodes have degree $K$ (see Section \ref{ProblemDefinition} for a formal definition). Otherwise, a TG (or LDPC code) is \emph{irregular} and degrees of the nodes can be expressed with a \emph{degree distribution}. 

Iterative decoding algorithms, which have low complexity and low decoding latency due to the sparsity property of parity--check matrix, have been developed on TG \cite{ZF05, CDE+05}. 
Iterative decoding algorithms decide on whether each code symbol is 0 or 1 by calculating probabilities for the code symbols to estimate the original information. 
The calculated probabilities are dependent on each other if there are cycles on the TG. 
In order to minimize code symbol estimation errors, designing LDPC codes to maximize the smallest cycle length, i.e., $girth$, is useful. There are different approaches in the literature for obtaining a TG with large girth. 

One approach is to eliminate the cycles with length smaller than the target girth from a given TG. In \cite{MW03}, certain edges are exchanged within TG to eliminate small cycles without simultaneously creating any others. In the edge deletion algorithm of \cite{BTC+11}, an edge that is common for the maximum number of cycles is selected. These methods are heuristic approaches and they change the degree distribution of the nodes in the TG. It is known that the degree distribution affects the error correction capability of an LDPC code \cite{S09}. Hence, it is important to eliminate as few edges from TG as possible. There are studies based on optimization techniques in the literature to find the best degree distribution of an irregular TG in terms of error correction capability \cite{S09, PB15}. 


Another way of designing an LDPC code is to construct a TG from scratch. Bit--Filling heuristic in \cite{CM01} starts with a large girth target and decreases target as it inserts edges to TG one--by--one. The heuristic terminates when a prescribed girth is met. A randomized approach in \cite{DSB05} can create irregular LDPC codes by introducing new edges in a zig--zag pattern. Progressive Edge Growth (PEG) heuristic in \cite{HEA05} is based on adding edges to the TG iteratively without constructing small cycles. PEG algorithm is adjusted to generate a regular LDPC code in \cite{CC07} and an irregular LDPC code in \cite{HL12} for improving the error correction performance. Independent tree--based heuristic of \cite{PP11} can iteratively construct regular LDPC codes whose girth values are better than the ones obtained by PEG. A protograph is a TG with a relatively small number of nodes.
Design of LDPC codes with simple protographs is investigated in \cite{DDJ05} to obtain infinite dimensional LDPC codes. Different studies in the literature focus on the design of LDPC codes with large girth using the protograph \cite{EHB09, PST13}.  

Algebraic construction is to construct structured LDPC with algebraic and combinatorial methods. Turbo LDPC (T--LDPC) codes are structured regular codes whose TG includes two trees connected by an interleaver. In \cite{LM07}, authors design the interleaver to avoid small cycles and obtain T--LDPC codes with high girth. Quasi--cyclic LDPC (QC--LDPC) codes consist of identity matrices whose columns are shifted by a certain amount. A method that can build QC--LDPC codes with girth at least 6 using Vandermonde matrices is introduced in \cite{BCH08}. A technique to generate irregular QC--LDPC codes with girth at least 8 is given in \cite{ZBQ+14}. Quasi--cycle constraints are added to PEG algorithm in order to obtain regular and irregular QC--LDPC codes in \cite{LK04}. Other studies also use PEG algorithm for this code family \cite{PPS11} -- \cite{JLW+16}. For the same code family, a lifting method is given in \cite{MYK06} and generalized polygones are used in \cite{LP05}. Patent \cite{YW13} describes a method for QC--LDPC codes, that guarantees a girth of at least 8. 











The above mentioned methods are heuristic approaches and they may fail to generate a TG for a given dimension with a target girth value. On the other hand, optimization techniques are capable of finding a TG for a given girth value, or proving that there cannot be such a TG. Combinatorial approaches to design QC--LDPC codes are utilized in \cite{BKJ13} to find the best degree distribution of the nodes in a TG.  Authors obtain the degree distribution by evaluating all alternatives with respect to some performance metrics and choosing the most promising one. Then, authors construct a TG for the selected degree distribution. In \cite{HZD+15}, the selection criteria of PEG algorithm to locate an edge in a TG is modified in order to have a better girth value than PEG. 
The generated TG does not necessarily have the largest girth value, since their method is a TG constructive heuristic. There are other LDPC code constructive heuristics in the literature that avoid small cycles \cite{BK16} -- \cite{JHW+17}.  A genetic algorithm to design a TG with a small number of nodes is given in \cite{BDG+16}. In \cite{SSG14} a modified shortest--path algorithm is used to construct a TG.

Our contribution to the literature can be listed as follows:

\begin{itemize}
	\item We investigate the LDPC code design problem, which seeks a TG of desired dimension with a target girth value, from an optimization point of view.

	\item We propose an integer programming formulation to generate LDPC codes with a given girth value and develop a branch--and--cut algorithm for its solution. 

	\item  We investigate structural properties of the problem  for $(J, K)-$regular codes to improve our algorithm by applying a variable fixing scheme, adding valid inequalities and utilizing an initial solution generation heuristic. Our computational results indicate that our proposed methods significantly improve solvability of the problem.  


	\item We also illustrate how our method can be used to find the smallest dimension $n$ that one can generate a $(J, K)-$regular code (see Table \ref{tab:LBUBonN}).
\end{itemize}


The remainder of the paper is organized as follows: we formally define the problem and introduce our mathematical formulation in the next section. Section \ref{SolutionMethods} explains the proposed branch--and--cut method and techniques to improve its performance. We test the efficiacy of our methods via computational experiments in Section \ref{ComputationalResults}. Some concluding remarks and comments on future work appear in Section \ref{Conclusions}. 

\section{Problem Definition}\label{ProblemDefinition}

\begin{figure}[h]
\begin{center}
	\includegraphics[width=0.55\columnwidth]{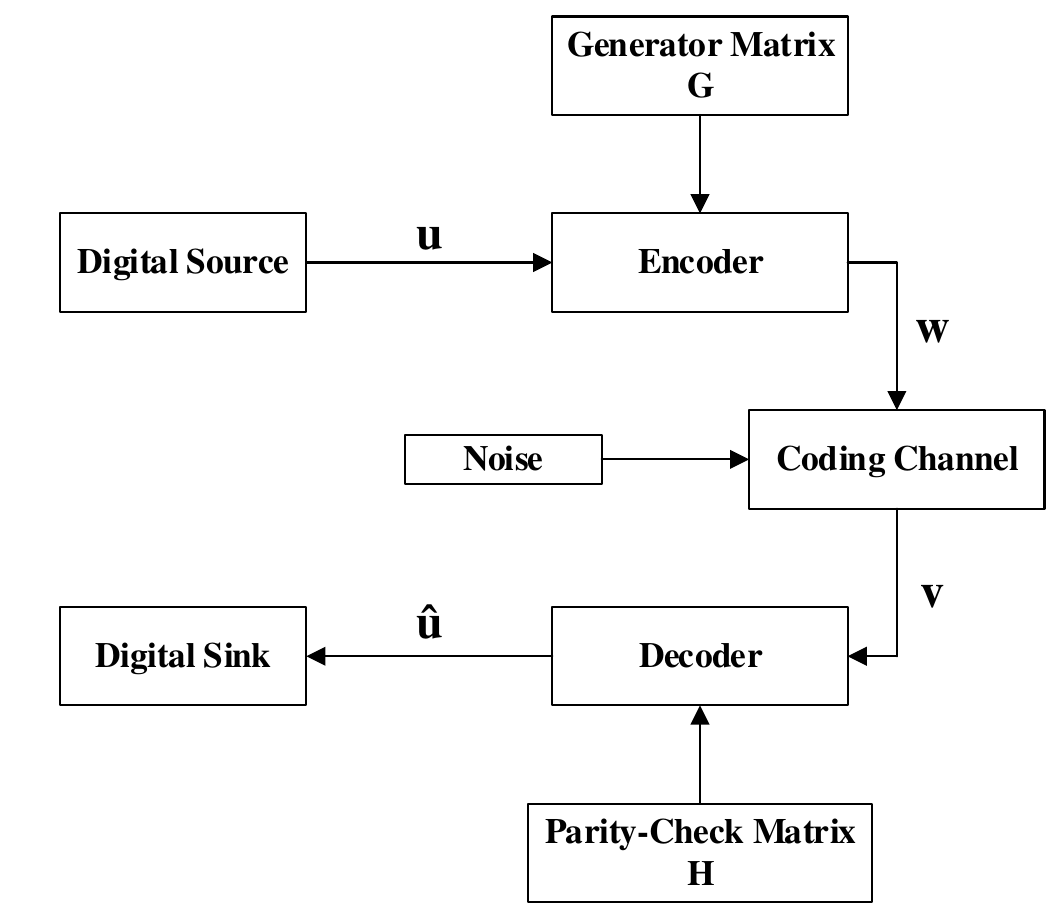}
\end{center}
\caption{Digital communication system diagram}
\label{fig:diagram}
\end{figure}

Figure \ref{fig:diagram} shows information flow in a digital communication system. In Figure \ref{fig:diagram}, let the original information be a binary vector $\mathbf{u} = (u_{1}u_{2}... u_{k})$ of $k$--bits, i.e.,  $u_{i} \in \{0,  1\}$. Encoder adds redundant parity--check bits to vector $\mathbf{u}$ by utilizing a $k \times n$ generator matrix $\mathbf{G}$.  That is  codeword  $\mathbf{w} =(w_{1}w_{2}... w_{n})$ of $n$--bits, where $n \geq k$ and $w_{i} \in \{0,  1\}$, is obtained through the operation  $\mathbf{w} = \mathbf{uG}$. In a codeword  $\mathbf{w}$, there are $k$ information bits and $(n - k)$ parity--check bits, which are used to test whether there are errors in the transmission. For integrity of the communication,  codeword  $\mathbf{w}$  should be in the null space of the $(n-k) \times n$   parity--check matrix $\mathbf{H}$, i.e., $\mathbf{w}\mathbf{H}^\textrm{T}=\mathbf{0}$ (mod 2) holds.

After transmission, the receiver gets vector $\mathbf{v}$ of $n$--bits as shown in Figure \ref{fig:diagram}. Decoder detects whether the received vector $\mathbf{v}$ includes errors or not by checking whether the expression $\mathbf{v}\mathbf{H}^\textrm{T}$ is equal to vector $\mathbf{0}$ in (mod 2) or not. In the case that $\mathbf{v}$ is erroneous, the decoder attempts to determine error locations and fix them \cite{L05}. As a result, the information $\mathbf{u}$ sent from the source is estimated as $\hat{\mathbf{u}}$ at the sink.

\begin{figure}[h]
\begin{center}
	\includegraphics[width=0.3\columnwidth]{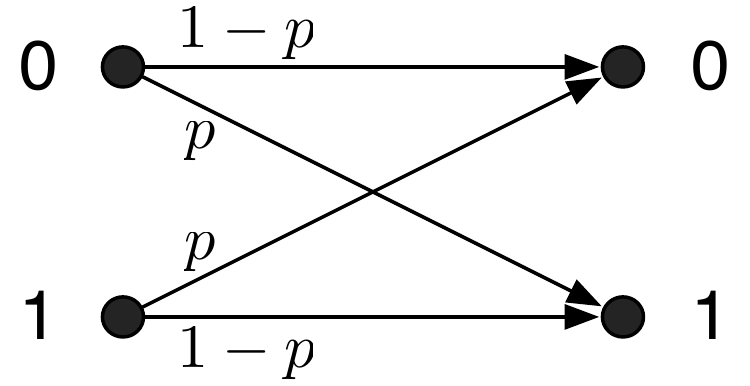}\\
\end{center}
\caption{Binary symmetric channel }
\label{fig:BSC}
\end{figure}

In this work, we focus on the binary symmetric channel (BSC) for modeling the noisy communication channel. As shown in Figure \ref{fig:BSC}, in a BSC, an error occurs with probability $p$ and the transmitted bit flips, i.e., if a bit is 0, it becomes 1 and vice versa. The transmission is completed without any errors with probability $1-p$  \cite{S03}. The decoder aims to find the locations of the errors in BSC. Once the decoder detects a bit is erroneous, it corrects the error by flipping the bit's value.


LDPC codes are members of linear block codes that can be represented by a sparse parity--check matrix $\mathbf{H}$, i.e., the number of ones at every row and column of the $\mathbf{H}$ matrix is forced to be very small. An LDPC code is \emph{regular}, if there are constant number of ones at each column and row of the matrix. As given in Figure \ref{fig:HMatrix}, a $(3, 6)-$regular LDPC code has only 3 ones at each column and 6 ones at each row independent from the dimension of the $\mathbf{H}$. This implies that for  $(3, 6)-$regular LDPC code with dimension $1500 \times 3000$, only  0.2\% of the matrix elements are nonzero. 


\vspace{-2mm}
\begin{figure}[h]
\begin{center}
\begin{equation*}
\mathbf{H}=\begin{bmatrix}
0 &1  &0  &0  &1  &1  &1  &1 &1 &0  \\
0 &0  &0  &0  &1  &1  &1  &1 &1 &1  \\
1 &0  &1  &1  &1  &1  &1  &0 &0 &0  \\
1 &1  &1  &1  &0  &0  &0  &0 &1 &1  \\
1 &1  &1  &1  &0  &0  &0  &1 &0 &1  \\
\end{bmatrix}
\end{equation*}
\end{center}
\caption{A parity--check matrix from $(3, 6)-$regular LDPC code family}
\label{fig:HMatrix}
\end{figure}

\begin{figure}[h]
\begin{center}
	\includegraphics[width=0.75\columnwidth]{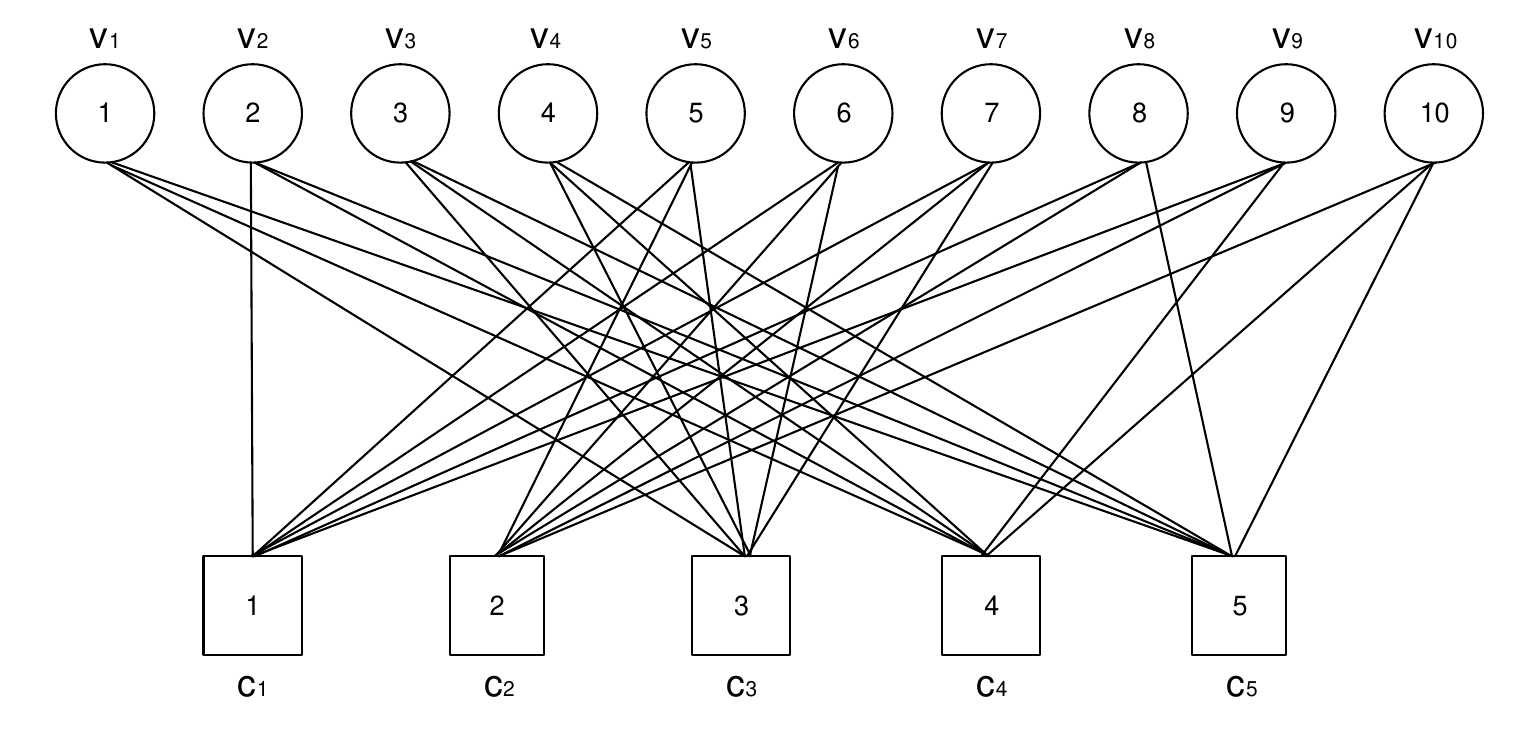} 
\end{center} \vspace{-2mm}
\caption{TG representation of the parity--check matrix given in Figure \ref{fig:HMatrix}}
\label{fig:TG}
\end{figure}

An LDPC code can alternatively be represented as a TG, which is a sparse bipartite graph, corresponding to the $\mathbf{H}$ matrix  \cite{T81}. On one part of the TG there is a variable node $j$ ($v_j$), $j \in \{1, ..., n\}$, for each bit of received vector. Each row of the $\mathbf{H}$ matrix represents a parity--check equation and corresponds to a check node $i$ ($c_i), i \in \{1, ..., n-k\}$, on the other part of the TG. A check node is said to be satisfied if its parity--check equation is equal to zero in (mod 2). The \emph{degree} of $v_j$ ($c_i$) is the number of adjacent check nodes (variable nodes) on the TG. Hence, $\mathbf{H}$ matrix is the bi--adjacency matrix of the TG. This representation of LDPC codes is practical due to the advantage of applying iterative decoding algorithms easily. Figure \ref{fig:TG} shows the TG representation of the $\mathbf{H}$ matrix defined in Figure \ref{fig:HMatrix}. 

\vspace{-2mm}

It is known that iterative decoding algorithms may fail to decode in the existance of small cycles (such as $(v_1, \ c_3, \ v_4, \ c_4)$ in Figure \ref{fig:TG}) \cite{R03}. The length of a smallest cycle is known as the $girth$ of the graph \cite{D10}. In this work, we will focus on designing LDPC codes whose TGs do not contain small cycles. 
In particular, we aim to construct a TG with girth no smaller than a given target girth value. 

\section{Solution Methods}\label{SolutionMethods}


In this section, we  introduce our integer programming formulations and propose a branch--and--cut algorithm for the solution of the problem. 
We investigate additional methods to improve the performance of our branch--and--cut algorithm. 
 We summarize the terminology used in this paper in Table \ref{tab:lop}.

\begin{onehalfspace}
\footnotesize
\begin{center}
\captionof{table}{List of symbols}
    \label{tab:lop}
\begin{tabular}{l l} 
\hline
\multicolumn{2}{c}{\textit{Parameters}} \\
\cline{1-2}
$k$    & length of the original information  \\
$n$  & length of the encoded information, number of columns in $\mathbf{H}$   \\
$m$ & $n-k$, number of rows in $\mathbf{H}$ \\
$\mathbf{G}$ & generator matrix \\
$\mathbf{H}$ & parity--check matrix \\
$p$  & error probability in BSC   \\
$T$ & target girth \\
$v_j$ & variable node $j$ \\
$c_i$ & check node $i$ \\
$dv_j$ & target degree of $v_j$ \\
$dc_i$ & target degree of $c_i$\\
$\rho(i,j)$ & cycle region of $(i,j)$ \\
\hline
\multicolumn{2}{c}{\textit{Decision Variables}} \\
\cline{1-2}
$X_{ij}$ & $(i,j)$ entry of the $\mathbf{H}$ matrix \\
$dv^s_j$ & slack for degree of $v_j$ \\
$dc^s_i$ & slack for degree of $c_i$ \\
\hline
\end{tabular}
\end{center}
\end{onehalfspace}

\subsection{Mathematical Formulations} \label{sec:MathematicalFormulations}


In our Girth Feasibility Model (GFM), our aim is to generate an $\mathbf{H}$ matrix of dimensions $(m, n)$, where $m = n- k$, with girth no smaller than a given value $T$. In the GFM model given below, $X_{ij}$ variable represents the  $(i,j)$ entry of the $\mathbf{H}$ matrix, $dv_j$ is the degree of variable node $j$, and $dc_i$ is the degree of check node $i$. Constraints (\ref{degree1}) and (\ref{degree2}) allow generation of an irregular code with the given degree values. As a special case,  one can obtain a $(J, K)-$regular $\mathbf{H}$ matrix by picking $dv_j = J$ for all $j$ and $dc_i = K$ for all $i$. 

We introduce cycle breaking constraints (\ref{cons2}) for the cycles with length less than the target girth $T$. 
In GFM, the objective is a constant, since the target girth $T$ is a given value. Hence, any feasible solution of the model will be optimal. 

\textbf{Girth Feasibility Model (GFM):}
\vspace{-10pt}
\begin{align}
\mbox{max} & \;\; T\\ 
\mbox{s.t.: } & \sum_{i =1}^{m}X_{ij} = dv_j, \; j = 1, ..., n  \label{degree1} \\
& \sum_{j =1}^{n}X_{ij} = dc_i, \; i = 1, ..., m  \label{degree2} \\
& \sum_{(i,j) \in C}X_{ij} \leq |C| -1, \;\forall C \; \mbox{\textit{cycle with }} |C|  < T \label{cons2} \\
& X_{ij} \in \{0, 1\},  \;\; i = 1, ..., m, \; j = 1, ..., n.  \label{xvar}
\end{align}


An alternative modeling approach is to assume $dv_j$ and $dc_i$ as the target degrees of $v_j$ and $c_i$, respectively. In Minimum Degree Deviation Model (MDD), the objective is to minimize the degree deviations $dv^s_j$ of  $v_j$ and $dc^s_i$ of  $c_i$ from the target values. 


\textbf{Minimum Degree Deviation Model (MDD):}
\vspace{-10pt}
\begin{align}
\mbox{min} & \;\;  \sum_{j =1}^{n} dv^s_j  + \sum_{i =1}^{m} dc^s_i  \\ 
\mbox{s.t.: } & \sum_{i =1}^{m}X_{ij} +  dv^s_j = dv_j, \; j = 1, ..., n  \label{degree3} \\
& \sum_{j =1}^{n}X_{ij} + dc^s_i = dc_i, \; i = 1, ..., m  \label{degree4} \\
& (\ref{cons2}) - (\ref{xvar}) \\
& dv^s_j , dc^s_i \geq 0,  \;\; i = 1, ..., m, \; j = 1, ..., n. 
\end{align}

One can observe that MDD is always feasible, since $X_{ij} = 0$ for all $(i,j)$, $dv^s_j = dv_j$ for all $j$, and $dc^s_i = dc_i$ for all $i$ is a trivial solution. Moreover, if the optimum objective function value of MDD is zero, which means constraints (\ref{degree3}) and (\ref{degree4}) are satisfied without deviation, we get a feasible (optimum) solution of GFM.

As we explain in Proposition \ref{prop8}, GFM can be infeasible depending on the value of the target girth $T$.  Hence, in our study, we work with the MDD model. Since there can be an exponential number of cycles in a TG, we can have exponential number of constraints (\ref{cons2}) in the corresponding MDD model. In order to obtain a solution in an acceptable amount of time, we add the constraints (\ref{cons2}) in a cutting--plane fashion to MDD. This gives rise to our branch--and--cut algorithm explained in the next section. 


\subsection{Branch--and--Cut Algorithm} \label{sec:BranchandCutAlgorithm}

The main steps of our Branch--and--Cut (BC) algorithm are listed in Algorithm 1. In the BC algorithm, we are given a target girth value $T$ and the dimensions of $\mathbf{H}$ matrix as $(m, n)$. We initialize our algorithm by relaxing constraints (\ref{cons2}) from MDD, to obtain relaxed model $\text{MDD}^r$. Steps $(I.1) - (I.3)$ are our improvement techniques (see Section \ref{sec:Improvements}) to the BC algorithm.  

\vspace{-2mm}

\begin{onehalfspace}
\begin{center}
\footnotesize
$
\begin{tabular}{ll}
\textbf{Algorithm 1:} (Branch--and--Cut) \\
\hline
\vspace{-4mm}\\
\textbf{Input:} Target girth value $T$, $(m, n)$ \\ 
\hline
\vspace{-4mm}\\
0. Obtain $\text{MDD}^r$ by removing constraints (\ref{cons2}) from MDD, set $x^* = null$ and $z^* = \infty$.\\ 
$(I.1)$ Apply Algorithm 4 to fix some $X_{ij}$ variables, update $x^*$ and $z^*$. \\
$(I.2)$ Add valid inequalities given in Proposition \ref{prop3} to $\text{MDD}^r$. \\
$(I.3)$ Apply Algorithm 6 to generate a feasible solution, update $x^*$ and $z^*$. \\ 
 \hspace{10pt} add $\text{MDD}^r$ to list $\mathcal{L}$.\\
1. \textbf{While} list $\mathcal{L}$ is not empty \\
2. \hspace{10pt} Select and remove a problem from $\mathcal{L}$.\\
3. \hspace{10pt} Solve LP relaxation of the problem. \\
4. \hspace{10pt}  \textbf{If} the solution is infeasible,  \textbf{Then} prune the branch and go to Step 1. \\
5. \hspace{10pt}  \textbf{Else} let the current solution be $x$ with objective value $z$.\\
6. \hspace{10pt} \textbf{End If}\\
7. \hspace{10pt}  \textbf{If} $z \geq z^*$,  \textbf{Then} prune the branch and go to Step 1. \\ 
8. \hspace{10pt}  \textbf{If} $x$ is an integer solution,  \\
 \hspace{39pt} \textbf{If} Algorithm 2 finds cycles smaller than $T$, \textbf{Then} add cuts (\ref{cons2}) and go to Step 3. \\
  \hspace{39pt} \textbf{Else} set $z^* \leftarrow z$, $x^* \leftarrow x$. \\ 
 \hspace{39pt} \textbf{End If}\\
9. \hspace{10pt}  \textbf{Else If} Algorithm 3 generates any cuts, \textbf{Then} add cuts (\ref{cons2}) and go to Step 3.\\
10.  \hspace{5pt}  \textbf{Else} branch to partition the problem into subproblems. \\
 \hspace{50pt} Add these problems to $\mathcal{L}$ and go to Step 1. \\
11. \hspace{5pt} \textbf{End If}\\
12. \textbf{End While}\\
\hline
\vspace{-4mm}\\
\textbf{Output:} $\mathbf{H}$ matrix with girth at least $T$ \\
\hline
\end{tabular}
$
\end{center}
\end{onehalfspace}
\vspace{5mm}

We can find either an integral or a fractional solution after solving the relaxed MDD. In the case we find an integral solution, we test its feasibility with respect to the relaxed constraints (\ref{cons2}) with Algorithm 2. The integral solution is separated from the solution space by adding required constraints from (\ref{cons2}) if the solution is not feasible. Similarly, we try to separate a fractional solution from the solution space with Algorithm 3, in order to strengthen the linear relaxation of MDD.




\vspace{-3mm}
\begin{figure}[!h]
\begin{center}
	\includegraphics[width=0.35\columnwidth]{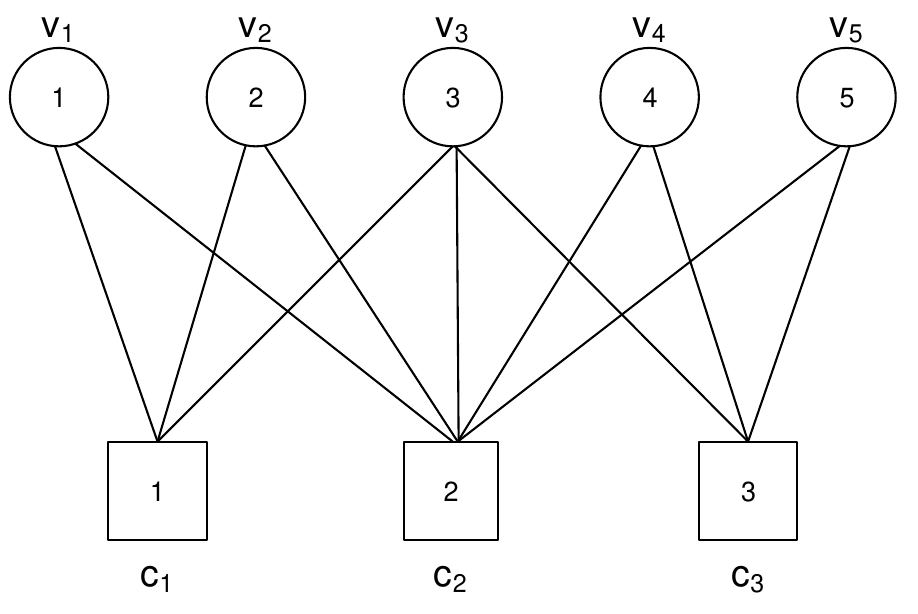}
\end{center} \vspace{-3mm}
\caption{An example TG for Algorithm 2}
\label{fig:TG2}
\end{figure}

\begin{figure}[h]
\begin{center}
	\includegraphics[width=1.05\columnwidth]{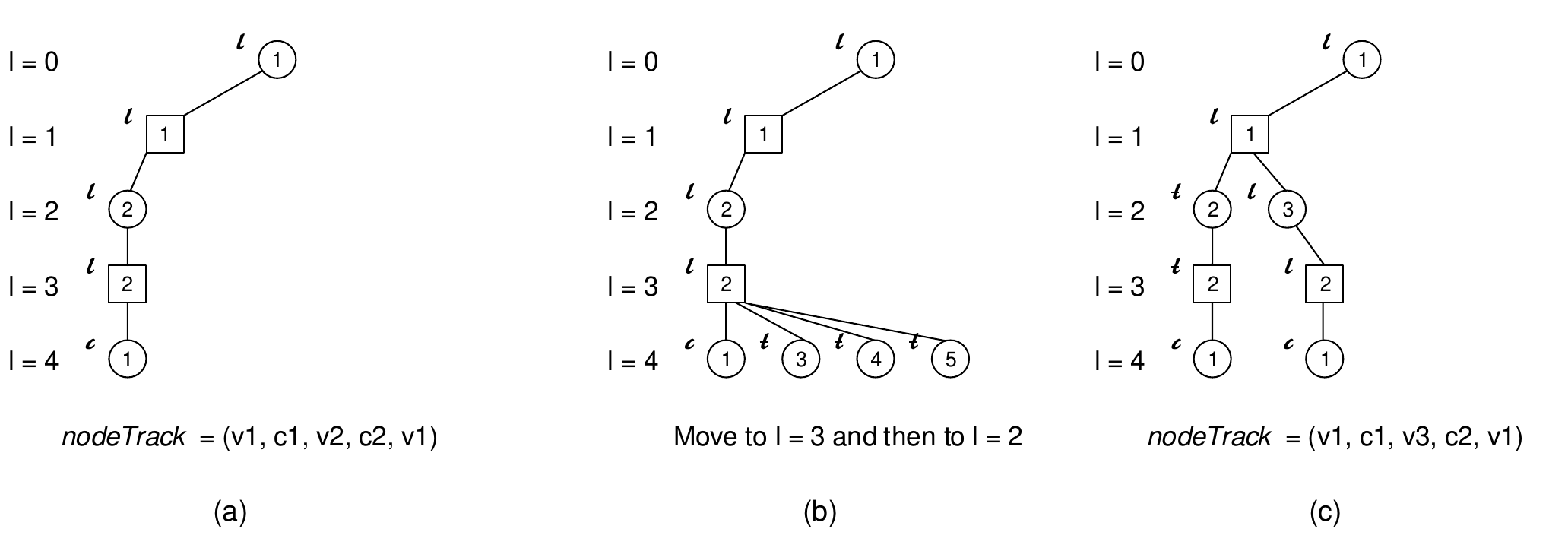}
\end{center} 
\caption{Depth--first--search in integral solution separation}
\label{fig:DFS}
\end{figure}

In the integral solution separation problem, we find all cycles in the TG whose length is less than $T$ with a depth--first--search algorithm running in $\mathcal{O}(|V| + |E|)$ time using Algorithm 2. In Figure \ref{fig:DFS}, we illustrate Algorithm 2 with $T = 6$ on the TG given in Figure \ref{fig:TG2}. In Figure \ref{fig:DFS}a, the search algorithm starts with $v_1$ at level 0, i.e., $l = 0$, and it is labeled. We label $c_1$ at $l = 1$,  $v_2$ at $l = 2$ and  $c_2$ at $l = 3$, since they are the first untracked neighbors of their predecessors. At $l = 4$, we visit $v_1$ but it has been previously labeled. This means that we have a cycle of length--4 consisting of nodes stored in $nodeTrack$ array and we add this cycle to $\mathcal{C}$ set, which keeps all cycles whose length is less than $T$ in the current integral $\mathbf{H}$ matrix. 


\begin{onehalfspace}
\begin{center}
\footnotesize
$
\begin{tabular}{ll}
\textbf{Algorithm 2:} (Integral Solution Separation) \\
\hline
\vspace{-4mm}\\
\textbf{Input:} A solution of $\text{MDD}^r$ with integral $X_{ij}$ values, $T$ target girth \\
\hline
\vspace{-4mm}\\
1. Let set of cycles $\mathcal{C} = \emptyset$ and $nodeTrack$ be an array \\
2. \textbf{For Each} variable node $j$, let $l = 1$\\
3. \hspace{10pt}  \textbf{While} $l > 0$, \textbf{Do} 
set $nodeTrack[0] = j$ and label node $j$ \\
4. \hspace{20pt} \textbf{For Each} level $l$ from 1 to $T-2$ \\
5. \hspace{30pt} Set $nodeTrack[l]$ to first untracked neighbor of  $nodeTrack[l-1]$\\ 
6. \hspace{30pt} \textbf{If}  $nodeTrack[l]$ is labeled, \textbf{Then} a cycle of length $l$ is added to  $\mathcal{C}$ \\
 \hspace{60pt} unlabel $nodeTrack[l]$ and \\ 
 \hspace{60pt} go to next untracked neighbor of  $nodeTrack[l-1]$\\  
\hspace{60pt}  \textbf{If} no such neighbor, \textbf{Then}  $l \leftarrow l-1$\\
7. \hspace{30pt} \textbf{Else} label  $nodeTrack[l]$ and $l \leftarrow l+1$, \textbf{End If} \\
8. \hspace{20pt}  \textbf{For Each}\\
9. \hspace{10pt} \textbf{End While} \\
10. \textbf{End For Each} \\
\hline
\vspace{-4mm}\\
\textbf{Output:} Set of cycles $\mathcal{C}$ \\
\hline
\end{tabular}
$
\end{center}
\end{onehalfspace}
\vspace{5mm}

In Figure \ref{fig:DFS}b, we consider other untracked neighbors of $c_2$ at level 4. After observing that none of $v_3$, $v_4$ and $v_5$ form a cycle, we unlabel them and return to level 3. At $l = 3$, we see that there are no other untracked neighbors of $c_2$ and backtrack to level 2. In Figure \ref{fig:DFS}c, we see $v_3$ is untracked and we label it at $l = 2$. We label $c_2$ at $l = 3$ and $v_1$ at $l = 4$. This means we found another cycle of length--4 and add this to set $\mathcal{C}$. 

\vspace{-0.5mm}



The time to find an optimal solution of MDD can be improved by reducing the feasible region using cuts for fractional solutions. In such a case, we have fractional  $X_{ij}$ values in the TG. We consider finding a maximum average cost cycle in the TG with $X_{ij}$ as cost values. If this cycle violates constraints (\ref{cons2}) and its length is less than $T$, then we can add the corresponding violated constraint. 

Minimum mean cost cycle is a well known network problem in the literature and there is a polynomial time solution algorithm for the directed graphs \cite{AMO93}. The problem simply aims to find a directed cycle $C$ with the smallest mean cost  $\sum_{(i,j) \in C}X_{ij} / |C|$   in a graph. However, we cannot implement this algorithm directly, since a TG is undirected. 
For the solution, we can update best known mean cost by implementing a negative cycle detection algorithm repeatedly. Bellman--Ford algorithm can detect negative cycles while searching 1--to--many shortest paths for directed graphs. Bellman--Ford algorithm is also applicable for undirected graphs in $\mathcal{O}(|V||E|)$ time, if for an edge $(i,j)$ the algorithm updates distance label of node $j$ when it is not the predecessor of node $i$ \cite{AMO93}. If the algorithm detects a negative cycle, we can track the predeccessor list to form the cycle. 

In the fractional solution separation problem, we use the  undirected Bellman--Ford algorithm to detect negative cycles within a mean cost update method.  We first set edge costs as $-X_{ij}$ to turn our maximization problem to minimization. Let $\mu$ represent an estimation on the minimum mean cost, and $\mu^*$ denote the (unknown) optimal value of $\mu$. Then, given a $\mu$ value, we update the edge costs to $(-X_{ij} - \mu)$ and check for the existance of a negative cycle. If we start with a $\mu$ that is an upper bound for $\mu^*$, we can face with one of these cases  for the minimum mean cost $\mu^*$.

\textbf{Case 1:} $G$ has a negative cycle $C$. In this case, $\sum_{(i,j) \in C} (-X_{ij} - \mu) < 0$. This means,

\begin{equation}
\mu > -\frac{\sum_{(i,j) \in C} X_{ij}}{|C|} > \mu^*.
\end{equation}

Hence, $\mu$ is a strict upper bound on $\mu^*$. We can update  $\mu$ as $\mu = -\frac{\sum_{(i,j) \in C} X_{ij}}{|C|}$ in the next iteration.

\textbf{Case 2:} $G$ has a zero--cost cycle $C^*$. In this case, $\sum_{(i,j) \in C^*} (-X_{ij} - \mu) = 0$. This means,

\begin{equation}
\mu = -\frac{\sum_{(i,j) \in C^*} X_{ij}}{|C^*|} = \mu^*.
\end{equation}

Hence, $\mu = \mu^*$ and  $C^*$ is a minimum mean cost cycle. 

\vspace{3mm}

\begin{onehalfspace}
\begin{center}
\footnotesize
$
\begin{tabular}{ll}
\textbf{Algorithm 3:} (Fractional Solution Separation) \\
\hline
\vspace{-4mm}\\
\textbf{Input:} A solution of $\text{MDD}^r$ with fractional $X_{ij}$ values, $T$ target girth \\
\hline
\vspace{-4mm}\\
1. Let $\mu =0$, set cost of edge $(i,j)$ as $(-X_{ij} - \mu)$ \\
2. \textbf{While} we can detect negative cycle $C$ with undirected Bellman--Ford \\
3. \hspace{10pt} \textbf{If}  $|C| < T$ and $C$ is violating (\ref{cons2}), \textbf{Then} add corresponding cut (\ref{cons2}) \\
4. \hspace{10pt} Update $\mu \leftarrow -\frac{\sum_{(i,j) \in C} X_{ij}}{|C|}$ \\
5. \textbf{End While}\\
\hline
\vspace{-4mm}\\
\textbf{Output:} Cuts added to $\text{MDD}^r$ model \\
\hline
\end{tabular}
$
\end{center}
\end{onehalfspace}
\vspace{5mm}

Fractional solution separation algorithm is summarized in Algorithm 3. We set initial $\mu = 0$, since it is an upper bound on $\mu^*$. If we can find a negative cycle with length $|C| < T$, we can add a cut to MDD if it is violated. This means that $C$ is a cycle with $\sum_{(i,j) \in C} X_{ij} > |C| -1$. We continue updating $\mu$ values until we find a minimum mean cycle. 



\subsection{Improvements to the Branch--and--Cut Algorithm} \label{sec:Improvements}

In this section we propose some improvements to the BC algorithm given in the previous section. 
We first observe that the solution space of MDD includes symmetric solutions. Hence, we consider a variable fixing approach to decrease the adverse effect of symmetry. Secondly, we introduce some valid inequalities to improve the linear relaxation of MDD. Finally, we adapt an algorithm from the telecommunications literature, i.e., PEG, to provide an initial solution to the BC algorithm.

\subsubsection{Symmetry in the MDD Solution Space} \label{Symmetry}

\begin{figure}[h]
\begin{center}
	\includegraphics[width=0.7\columnwidth]{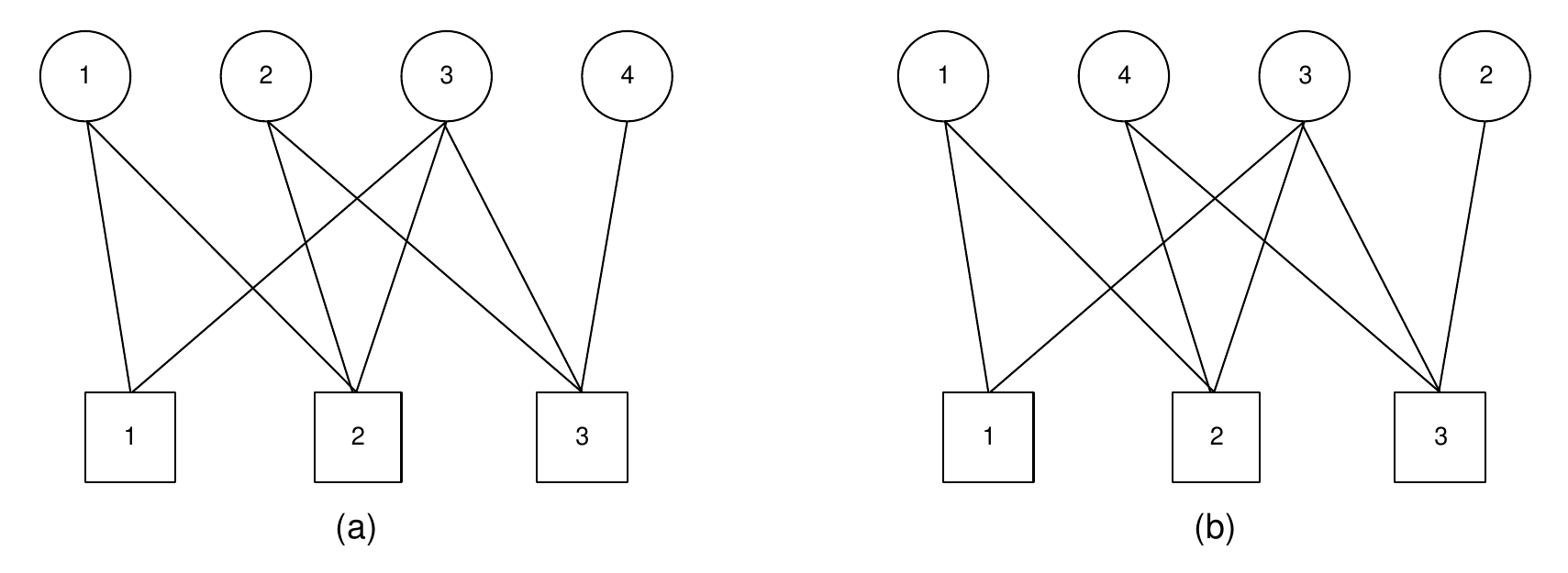}
\end{center} 
\caption{Symmetry in MDD solution space}
\label{fig:Symmetry}
\end{figure}

In combinatorial optimization problems such as scheduling, symmetry among the solutions is an important issue, which directly affects the performance of  applied solution methods \cite{JD13, SS01}. We observe that the feasible region of MDD contains symmetric solutions. That is, there can be isomorphic representations of a TG by permuting the variable and check nodes. As an example, the variable nodes are in the order of $\{v_1, v_2, v_3, v_4\}$ in Figure \ref{fig:Symmetry}a and the names of $v_2$ and $v_4$ are swapped in Figure \ref{fig:Symmetry}b. 


In Figure \ref{fig:HSymmetry}, $\mathbf{H_1}$ and $\mathbf{H_2}$ are the parity--check matrices for TGs in Figures \ref{fig:Symmetry}a and \ref{fig:Symmetry}b, respectively. We see that although TGs are isomorphic, their $\mathbf{H}$ matrix representations are not identical. In the MDD solution space $\mathbf{H_1}$ and $\mathbf{H_2}$ are considered as two different solutions, which increases the complexity of the solution algorithm.

\vspace{-6mm}
\begin{figure}[h]
\begin{center}
\begin{equation*}
\mathbf{H_1}=\begin{bmatrix}
1 &0  &1  &0    \\
1 &1  &1  &0   \\
0 &1  &1  &1   \\
\end{bmatrix} \hspace{10mm}
\mathbf{H_2}=\begin{bmatrix}
1 &0  &1  &0    \\
1 &0  &1  &1   \\
0 &1  &1  &1   \\
\end{bmatrix}
\end{equation*}
\end{center} \vspace{-3mm}
\caption{Parity--check matrices for the TGs in Figure \ref{fig:Symmetry}}
\label{fig:HSymmetry}
\end{figure}
\vspace{-3mm}

We can calculate the number of symmetric solutions for a TG as $(n!)(m!)$, since we can permute $n$ variable nodes as $(n!)$ and $m$ check nodes as $(m!)$ different ways.

\subsubsection{Symmetry Breaking with Variable Fixing} \label{sec:VariableFixing}

In the literature, ordering the decision variables, adding symmetry--breaking cuts to the formulation and reformulating the problem are some of the techniques to eliminate symmetric solutions from the feasible region \cite{SS01, XXKK17}. In our case, we propose a fixing scheme for nonzero $X_{ij}$ entries of $\mathbf{H}$ matrix that breaks symmetry and does not form any cycles in TG.  
 
\begin{figure}[!h]
\begin{center}
	\includegraphics[width=0.85\columnwidth]{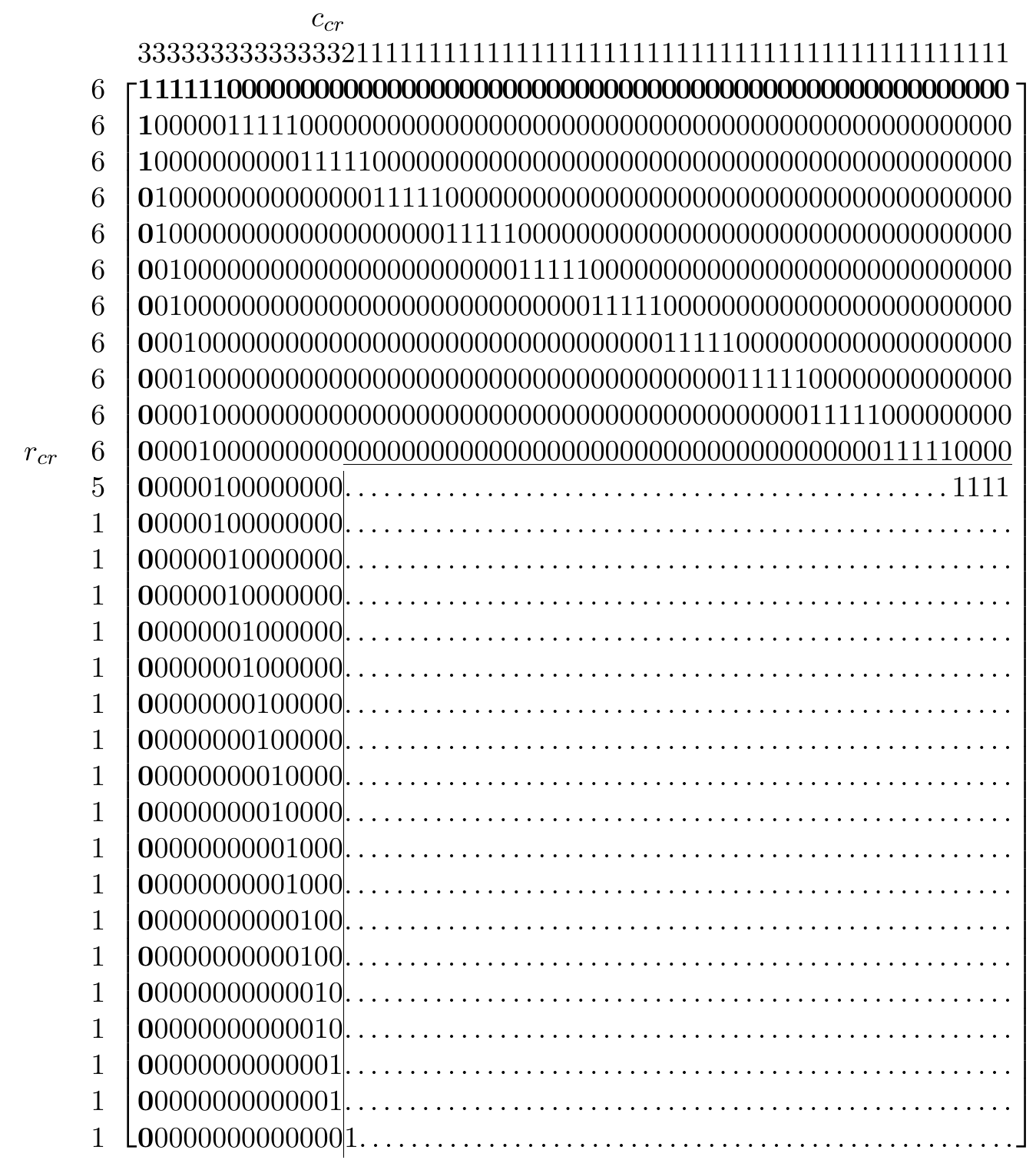}
\end{center} 
\caption{Variable fixing on a  $(3, 6)-$regular  $\mathbf{H}$ matrix of dimensions (30, 60)}
\label{fig:VariableFixing}
\end{figure}

In our variable fixing method (given as Algorithm 4) we consider $(J, K)$--regular $\mathbf{H}$ matrices and two modes, i.e., $basic$ and $extended$. 
 In the $basic$ mode, we fix first $K$ entries in the first row to 1 and first $J$ entries in the first column to 1. The remaining entries in the first row and column are set to 0, since constraints (\ref{degree1}) for $j = 1$ and constraints (\ref{degree2}) for $i = 1$ are satisfied. We illustrate the $basic$ and $extended$ modes in Figure \ref{fig:VariableFixing} for a $(3, 6)-$regular code of dimensions $(30, 60)$ below. Bold entries in Figure \ref{fig:VariableFixing} are fixed with the $basic$ mode.



\begin{onehalfspace}
\begin{center}
\footnotesize
$
\begin{tabular}{ll}
\textbf{Algorithm 4:} (Variable Fixing) \\
\hline
\vspace{-4mm}\\
\textbf{Input:} $(m, n)$ dimensions,  $(J, K)$ values,  $mode$ type \\
\hline
\vspace{-4mm}\\
0. Let $r_{cr} = \lfloor (n -1) / (K-1) \rfloor$ and $c_{cr} = \lfloor (m - 1) /(J-1) \rfloor$   \\
\hspace{10pt} Set  $X_{1j} = 0$, $j = 1, ..., n$, $X_{i1} = 0$,  $i = 1, ..., m$ \\
\hspace{10pt} \textbf{If} $mode = extended$ \\
\hspace{20pt} \textbf{For} $ i = 2, ..., r_{cr}, j = 1, ..., n$,  set $X_{ij} = 0$ \\
\hspace{20pt} \textbf{For} $ i = r_{cr} + 1, ..., m, j = 2, ..., c_{cr}$,  set $X_{ij} = 0$ \\
\hspace{10pt} \textbf{End If}\\
1.  Set  $X_{1j} = 1, j= 1, ..., K$ and $X_{i1} = 1, i= 1, ..., J$ \\
2.  \textbf{If} $mode = extended$ \\
3. \hspace{10pt} \textbf{For} $ i = 2, ..., r_{cr} + 1, j = 1, ..., K-1$,\\
4. \hspace{20pt}  \textbf{If} $1 + (i-1)(K-1) + j \leq n$, \textbf{Then} set  $X_{i, 1 + (i-1)(K-1) + j} = 1$. \\
5. \hspace{10pt} \textbf{End For}\\
6. \hspace{10pt} \textbf{For} $ i = 1, ..., J-1, j = 2, ..., c_{cr} + 1$,\\
7. \hspace{20pt}  \textbf{If} $1 + j(J -1) + i \leq m$, \textbf{Then} set  $X_{1 + j(J -1) + i, j} = 1$. \\
8. \hspace{10pt} \textbf{End For}\\
9.  \textbf{End If}\\
\hline
\vspace{-4mm}\\
\textbf{Output:} Some $X_{ij}$ values are fixed \\
\hline
\end{tabular}
$
\end{center}
\end{onehalfspace}
\vspace{5mm}

In the $extended$ mode, we extend variable fixing further as dimensions $(m, n)$ of the $\mathbf{H}$ matrix allow. 
In Figure \ref{fig:VariableFixing}, the labels on the rows and colums show the sum of the values in that row and column, respectively. We observe that for  $r_{cr} = \lfloor (n -1) / (K-1) \rfloor$ many rows the sum is equal to 6 and $c_{cr} = \lfloor (m - 1) /(J-1) \rfloor$ many columns the sum is equal to 3.
Hence, for $c_{cr}$--columns constraints (\ref{degree1}) and for $r_{cr}$--rows constraints (\ref{degree2}) are satisfied. We remain with a reduced rectangle of size $(m-r_{cr}) \times (n-c_{cr})$, which includes the unfixed $X_{ij}$ variables shown as dots. Algorithm 4 runs in $\mathcal{O}(nc_{cr})$ time.

In practical applications, for a $(J, K)-$regular code $J < K < n$ relationship is valid. In Proposition \ref{prop4}, we use this relationship to compare $r_{cr}$ and $c_{cr}$.

\begin{proposition}\label{prop4}
Let $J < K < n$. For a $(J, K)-$regular code of dimensions $(m, n)$,  $r_{cr} \leq c_{cr}$ where $r_{cr} = \lfloor (n -1) / (K-1) \rfloor$ and $c_{cr} = \lfloor (m - 1) /(J-1) \rfloor$. 
\end{proposition}

\emph{Proof.} Let $\frac{J}{K} = a \in (0,1)$, then $mK =nJ \implies m =na$. 
We can write, $\frac{m-1}{J-1} = \frac{na-1}{Ka-1} =\frac{a(n-1) + a -1}{a(K-1) + a -1}  >  \frac{n-1}{K-1}$, since $a < 1$. From here we obtain $ \lfloor\frac{n-1}{K-1}\rfloor \leq \lfloor\frac{m-1}{J-1}\rfloor \implies r_{cr} \leq c_{cr}$. $\square$\\

\begin{figure}[!h]
\begin{center}
	\includegraphics[width=0.4\columnwidth]{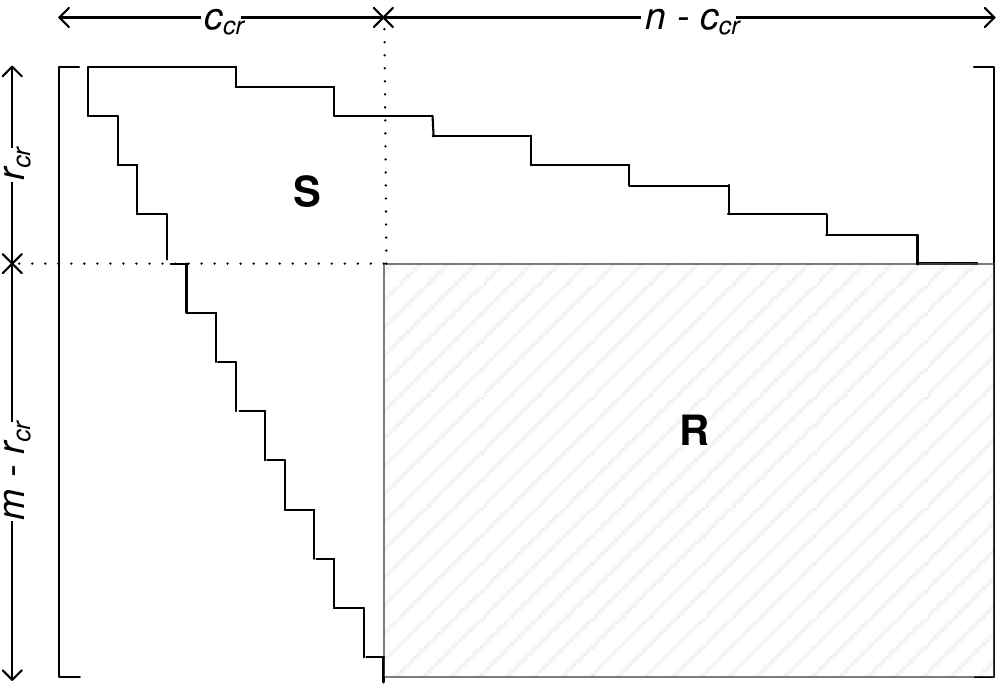}
\end{center} 
\caption{Reordered $(J, K)$--regular $\mathbf{H}$ matrix with girth  $T > \tau$}
\label{fig:Reorder}
\end{figure}

In Proposition \ref{prop5}, we show that any $(J, K)$--regular $\mathbf{H}$ matrix of dimensions $(m, n)$ that has sufficiently large girth $T$ can be expressed as in Figure \ref{fig:Reorder} by reordering its rows and columns.

\begin{proposition}\label{prop5}
Let $\mathbf{H}$ be a $(J, K)$--regular code of dimensions $(m, n)$. 
 Let $R$ be the reduced rectangle of size $(m-r_{cr}) \times (n-c_{cr})$ and $R \bigcup S$ be the region between the two extending 1--blocks as in Figure \ref{fig:Reorder}.
Let $\rho(i,j)$ be the length of a smallest cycle that is formed when $X_{ij} = 1$, and $\tau = \max_{(i,j) \in S}\{\rho(i,j)\}$.
 Then, nonzero entries of $\mathbf{H}$ can be represented as two extending 1--blocks as in Figure \ref{fig:Reorder} by reordering its rows and columns if it has a girth $T > \tau$. Remaining nonzero entries are in the reduced rectangle $R$.
\end{proposition}

\emph{Proof.} Let $\mathbf{H}$ be $(J, K)$--regular matrix of dimensions $(m, n)$ with girth $T > \tau$. Let us apply the following reordering algorithm with time complexity $\mathcal{O}(c_{cr})$ on the $\mathbf{H}$.

\vspace{3mm}

\begin{onehalfspace}
\begin{center}
\footnotesize
$
\begin{tabular}{ll}
\textbf{Algorithm 5:} (Reordering) \\
\hline
\vspace{-4mm}\\
\textbf{Input:} $\mathbf{H}$, $(m, n)$ dimensions,  $(J, K)$ values,  $T$ value \\
\hline
\vspace{-4mm}\\
1. Pick row 1, reorder columns such that all ones are in first $K$ columns.\\
\hspace{10pt}    Pick column 1, reorder rows such that all ones are in first $J$ rows. \\
2. \textbf{For} $s \in \{2, ..., r_{cr}\}$ \\ 
3. \hspace{10pt} Pick row $s$, reorder columns such that $(K-1)$ ones are in first available columns.\\
\hspace{25pt}    Pick column $s$, reorder rows such that $(J-1)$ ones are in first available rows. \\
4. \textbf{End For}\\
5. \textbf{For} $s \in \{ r_{cr}+ 1, ..., c_{cr}\}$ \\ 
6. \hspace{10pt} Pick column $s$, reorder rows such that $(J-1)$ ones are in first available rows. \\
7. \textbf{End For}\\
\hline
\vspace{-4mm}\\
\textbf{Output:} Reordered $\mathbf{H}$ matrix \\
\hline
\end{tabular}
$
\end{center}
\end{onehalfspace}
\vspace{5mm}

At step 1 of Algorithm 5, $J$ many ones are located in the first column. For the second row, i.e., $s = 2$, first available  $(K-1)$ columns to locate ones are the columns $(K +1, ..., 2K-1)$, since otherwise a cycle with length less than $T$ exists. Similarly for the second column, i.e., $s = 2$, first available $(J-1)$ rows are the rows $(J+1, ..., 2J-1)$ without creating a cycle. The algorithm continues in this fashion for $r_{cr}$ rows and columns. Since we see  in Proposition \ref{prop4} that $r_{cr} \leq c_{cr}$, we continue to locate ones for the remaining $(c_{cr} - r_{cr})$ many columns. $\square$\\

Using Proposition \ref{prop5}, we can give a lower bound on the dimension $n$ of a $(J, K)$--regular code with girth at least $T$ as in Proposition \ref{prop8}.

\begin{proposition}\label{prop8}
Consider a $(J, K)$--regular $\mathbf{H}$ matrix having girth at least $T$. Let $\rho(i,j)$ be the length of a smallest cycle that is formed when $X_{ij} = 1$.
The following statements are valid on dimensions $(m, n)$:
\begin{enumerate}
\item[(1)] $n = 2m$ if $K = 2J$,
\item[(2)] Consider Figure \ref{fig:Reorder} and let $(i, j) \in R \bigcup S$. Let $r_{cr}$ be the row such that $ \forall i \leq r_{cr}$ we have $\rho(i,j) < T$ and $\exists j$ with $\rho(r_{cr} + 1,j) \geq T$. Then 
\vspace{-5mm}
\begin{equation}
  n \geq (K-1)(r_{cr} + 1).
\end{equation}
\end{enumerate}
\vspace{-5mm}
\emph{Proof.} For a  $(J, 2J)$--regular $\mathbf{H}$ matrix, each variable node has $J$ neighbors and each check node has $2J$ neighbors in the TG. Since total variable degree should be equal to total check degree in a bipartite graph, we have $n J = m (2J) \implies n = 2m$. 

Since $\mathbf{H}$ is a $(J, K)$--regular matrix with girth at least $T$, we can reorder its rows and columns as in Figure \ref{fig:Reorder}. Let $(i, j)  \in R \bigcup S$. According to Proposition \ref{prop5}, the maximum dimension $n$ that this reordering is possible is such that $r_{cr} = \lfloor\frac{n-1}{K-1}\rfloor$ and $ \forall i \leq r_{cr}$ we have $\rho(i,j) < T$ and $\exists j \leq n$ with $\rho(r_{cr} + 1,j) \geq T$. From $r_{cr} = \lfloor\frac{n-1}{K-1}\rfloor$ we can write $r_{cr} + \frac{K - 2}{K-1} \leq \frac{n-1}{K-1}$ to maximize $n$. This gives $n \geq (K-1)(r_{cr} + 1)$. $\square$ 

\end{proposition}

We can calculate $\rho(i,j)$ of an entry $(i, j)$ by carrying out a breadth--first--search starting from the variable node $v_j$. The smallest depth which we revisit $v_j$ is $\rho(i,j)$. From Proposition \ref{prop8}, we can provide lower bound on $n$ for a (3,6)--regular code as $r_{cr} = 3, n \geq 20$ for $T = 6$, $r_{cr} = 13, n \geq 70$ for $T = 8$, $r_{cr} = 33, n \geq 170$ for $T = 10$ (see Figure \ref{fig:Subblocks} for $\rho(i,j)$ values).

\begin{figure}[!h]
\begin{center}
	\includegraphics[width=0.7\columnwidth]{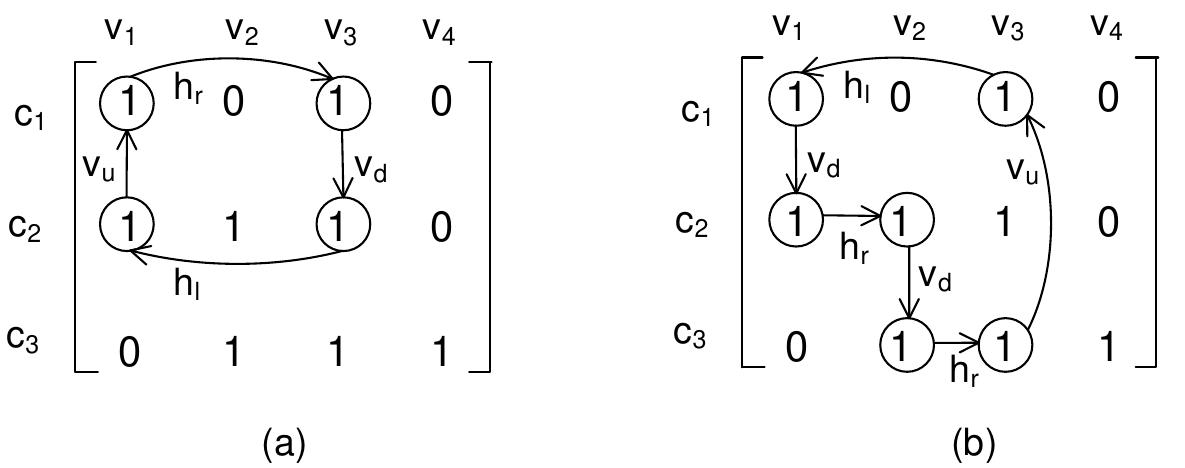} 
\end{center} 
\caption{Cycles $C_1$ and $C_2$ on $\mathbf{H_1}$ }
\label{fig:Cycles}
\end{figure}

Some characteristics of the cycles in a TG can be visualized by considering the TG given in Figure \ref{fig:Symmetry}a and the corresponding parity--check matrix $\mathbf{H_1}$ in Figure \ref{fig:HSymmetry}. It can be seen that $C_1 = (v_1, c_1, v_3, c_2)$ and $C_2 = (c_1, v_1, c_2, v_2, c_3, v_3)$ are two cycles in the TG in Figure \ref{fig:Symmetry}a. Figures \ref{fig:Cycles}a and \ref{fig:Cycles}b visualize cycles $C_1$ and $C_2$ on $\mathbf{H_1}$, respectively.  

We observe that a cycle is an alternating sequence of horizontal and vertical movements between cells having value 1. In particular, cycle $C_1$ is a sequence of horizontal right ($h_r$), vertical down ($v_d$), horizontal left  ($h_l$) and vertical up ($v_u$) movements. Similarly, cycle $C_2$ can be expressed with the sequence $(v_d, h_r, v_d, h_r, v_u, h_l)$. Moreover, we deduce that a cycle should include at least one from each of the $h_u$, $h_d$, $v_u$ and $v_d$ movements. 

\begin{proposition}\label{prop2}
Variable fixing on $\mathbf{H}$ matrix with the $extended$ mode does not form any cycles in the TG.
\end{proposition}

\begin{figure}[!h]
\begin{center}
	\includegraphics[width=0.8\columnwidth]{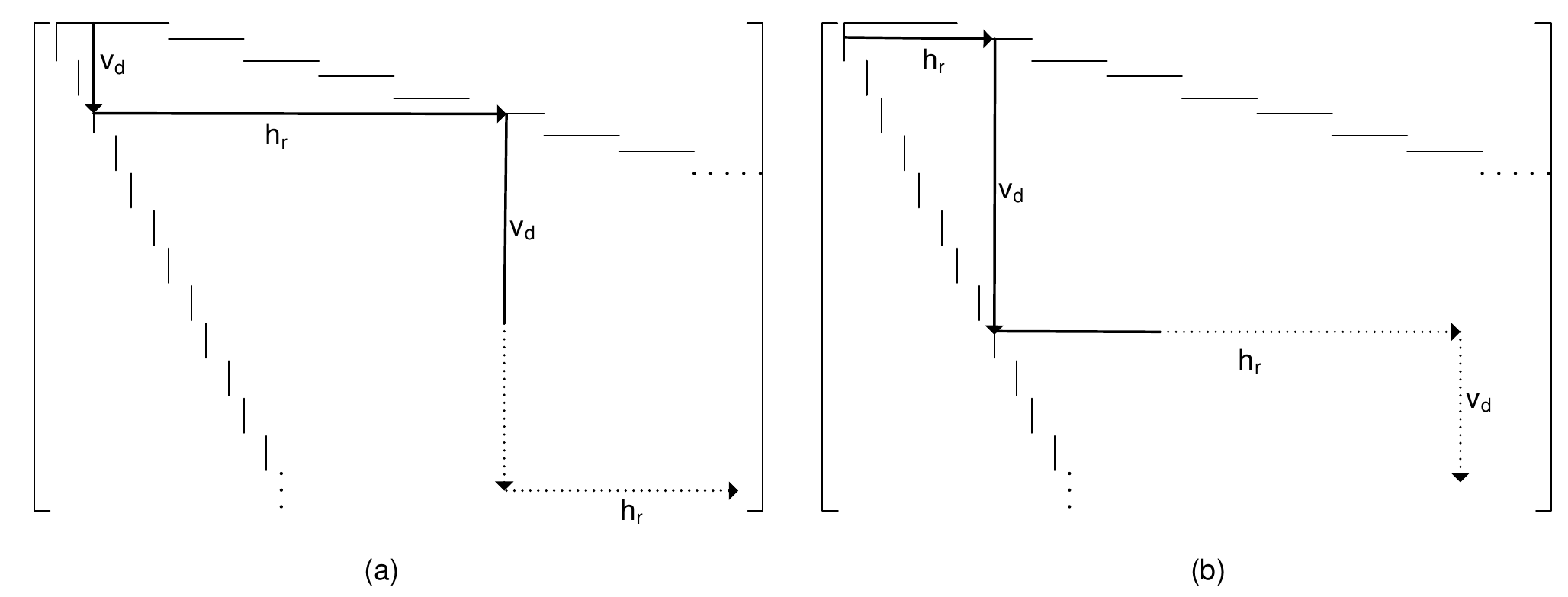}
\end{center} 
\caption{Alternating variable and check nodes, cases 1 and 2 }
\label{fig:CycleCase1}
\end{figure}

\begin{figure}[!h]
\begin{center}
	\includegraphics[width=0.8\columnwidth]{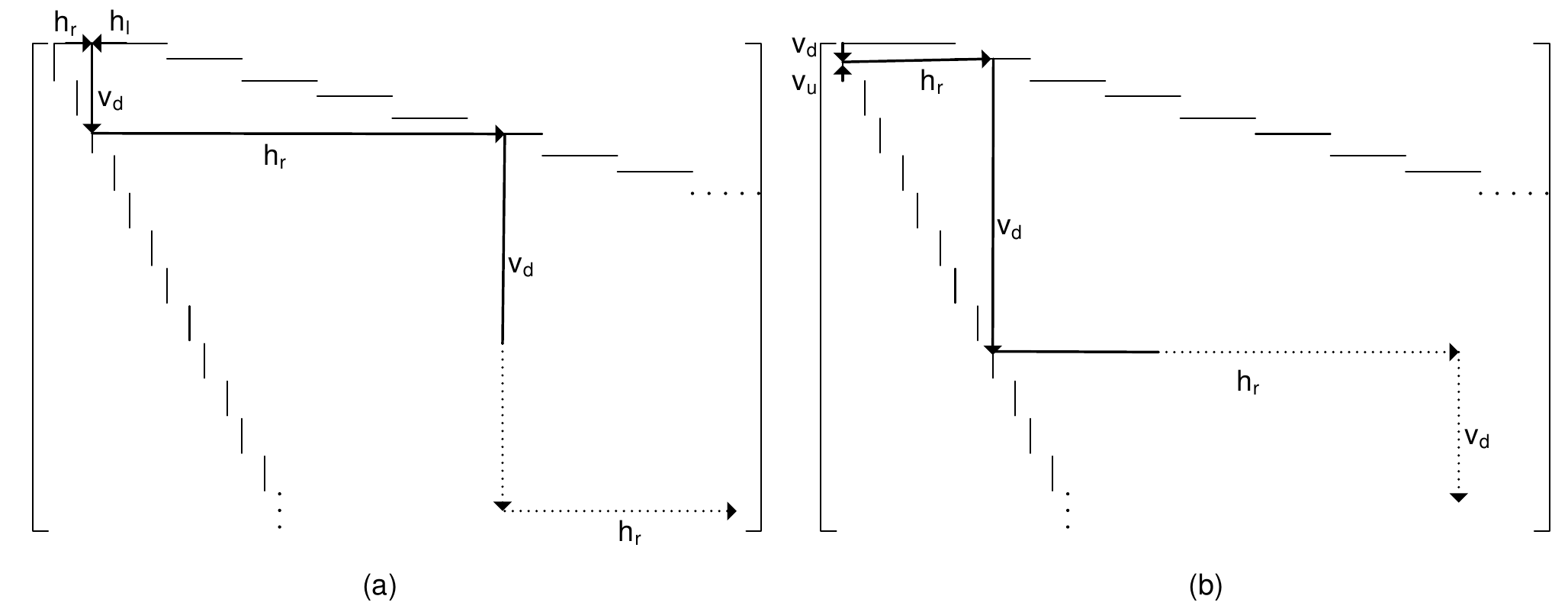} 
\end{center} 
\caption{Alternating variable and check nodes, cases 3 and 4 }
\label{fig:CycleCase2}
\end{figure}

\emph{Proof.} Assume we apply variable fixing with the $extended$ mode and consider cells whose $X_{ij}$ values have been fixed to 1. There are four cases to have an alternating sequence among variable and check nodes as given in Figures \ref{fig:CycleCase1} and \ref{fig:CycleCase2}. 

In Figure \ref{fig:CycleCase1}a, the sequence of case 1 is $(v_d, h_r, v_d, h_r, ...)$  and in Figure \ref{fig:CycleCase1}b for case 2, we have the sequence $(h_r, v_d, h_r, v_d, ...)$. Both of the sequences do not include $v_u$ and $h_l$ movements. Hence, there cannot be any cycles in these cases. 

In Figure \ref{fig:CycleCase2}a (case 3), we have two options to start, i.e., $h_r$ or $h_l$ movements. Then the sequence will be  $(h_r \text{ or } h_l, v_d, h_r, v_d, h_r, ...)$, which does not include $v_u$ movement. In Figure \ref{fig:CycleCase2}b (case 4), $v_d$ or $v_u$ are candidates to begin the sequence. In this case, the sequence will be  $(v_d \text{ or } v_u, h_r, v_d, h_r, v_d, ...)$, which does not include $h_l$ movement. Hence, there are no cycles in these cases either. $\square$\\

We can use the partial solution obtained with Algorithm 4 to generate a feasible solution of MDD. Since partial solution does not include any cycles (see Proposition \ref{prop2}), setting the nonfixed entries to zero gives a feasible solution (an upper bound). Step $(I.1)$ of Algorithm 1 implements variable fixing with the $basic$ or $extended$ mode and updates the initial upper bound.

\subsubsection{Valid Inequalities for Cycle Regions} \label{sec:ValidInequalities}


After applying extended fixing, MDD problem reduces to locating ones in the reduced rectangle $R$ of size $(m-r_{cr}) \times (n-c_{cr})$. That is problem size reduced by $\left(1 -\frac{(m-r_{cr}) \times (n-c_{cr})}{m \times n} \right)\times 100 \%$. We can further improve the performance of BC algorithm by introducing valid inequalities. 
We add the generated valid inequalities to MDD$^r$ at step $(I.2)$ of Algorithm 1. 

\begin{figure}[!h]
\begin{center}
	\includegraphics[width=1.1\columnwidth]{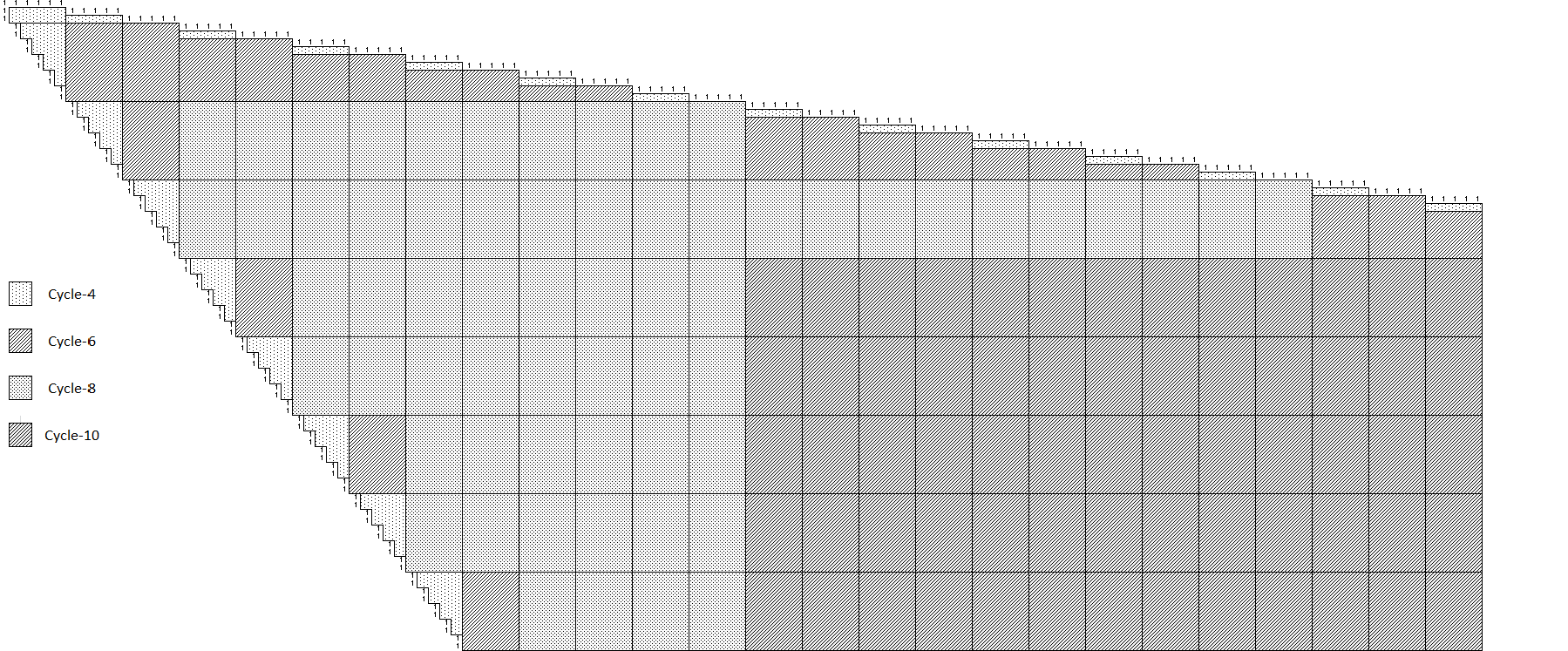}
\end{center} 
\caption{Subblocks and cycle regions with $J = 3$ and $K = 6$}
\label{fig:Subblocks}
\end{figure}

We observe that for given dimensions $(m, n)$, the reduced rectangle $R$ appears between the two extending 1--blocks as given in Figure \ref{fig:Reorder}. For a $(J, K)$--regular code, we divide the region $R \bigcup S$ into \emph{subblocks} with $(J- 1)(K - 1)$ rows and $(K - 1)$ columns as given in Figure \ref{fig:Subblocks}. For each entry $(i,j)$ in a subblock, we investigate the length of a smallest cycle $\rho(i,j)$ (see Proposition \ref{prop5}) when there is a single 1 at entry $(i,j)$. For example, in Figure \ref{fig:Subblocks}, we observe that  $\rho(i,j)$ is common for all $(i,j)$ entries in a subblock except the subblocks at the boundaries of the extending 1--blocks. Hence, we can define Cycle--4, Cycle--6, Cycle--8, and Cycle--10 regions, which have repeating pattern due to $(J, K)$--regularity.



\begin{figure}[!h]
\begin{center}
	\includegraphics[width=1\columnwidth]{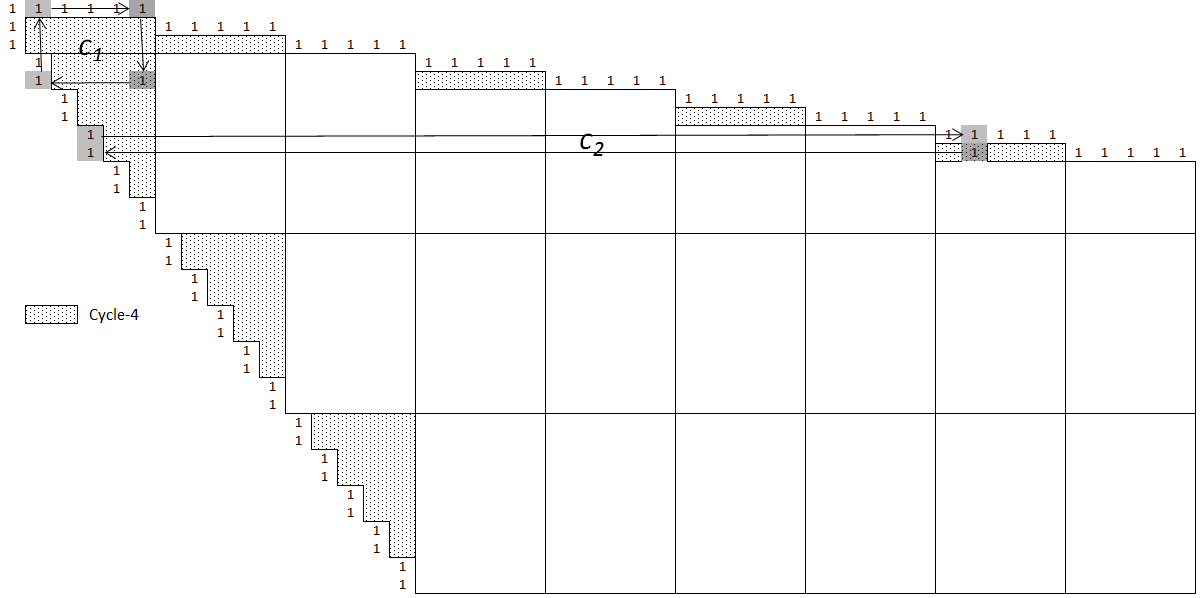}
\end{center} 
\caption{Cycle--4 regions with $J = 3$ and $K = 6$}
\label{fig:Cycle4}
\end{figure}

In particular, when there is a 1 in a Cycle--4 region (dotted area), we have a cycle of length 4 as in the case of cycles $C_1$ and $C_2$ in Figure \ref{fig:Cycle4}. We note that, Cycle--4 regions repeat both horizontally and vertically. 

\begin{figure}[!h]
\begin{center}
	\includegraphics[width=1.15\columnwidth]{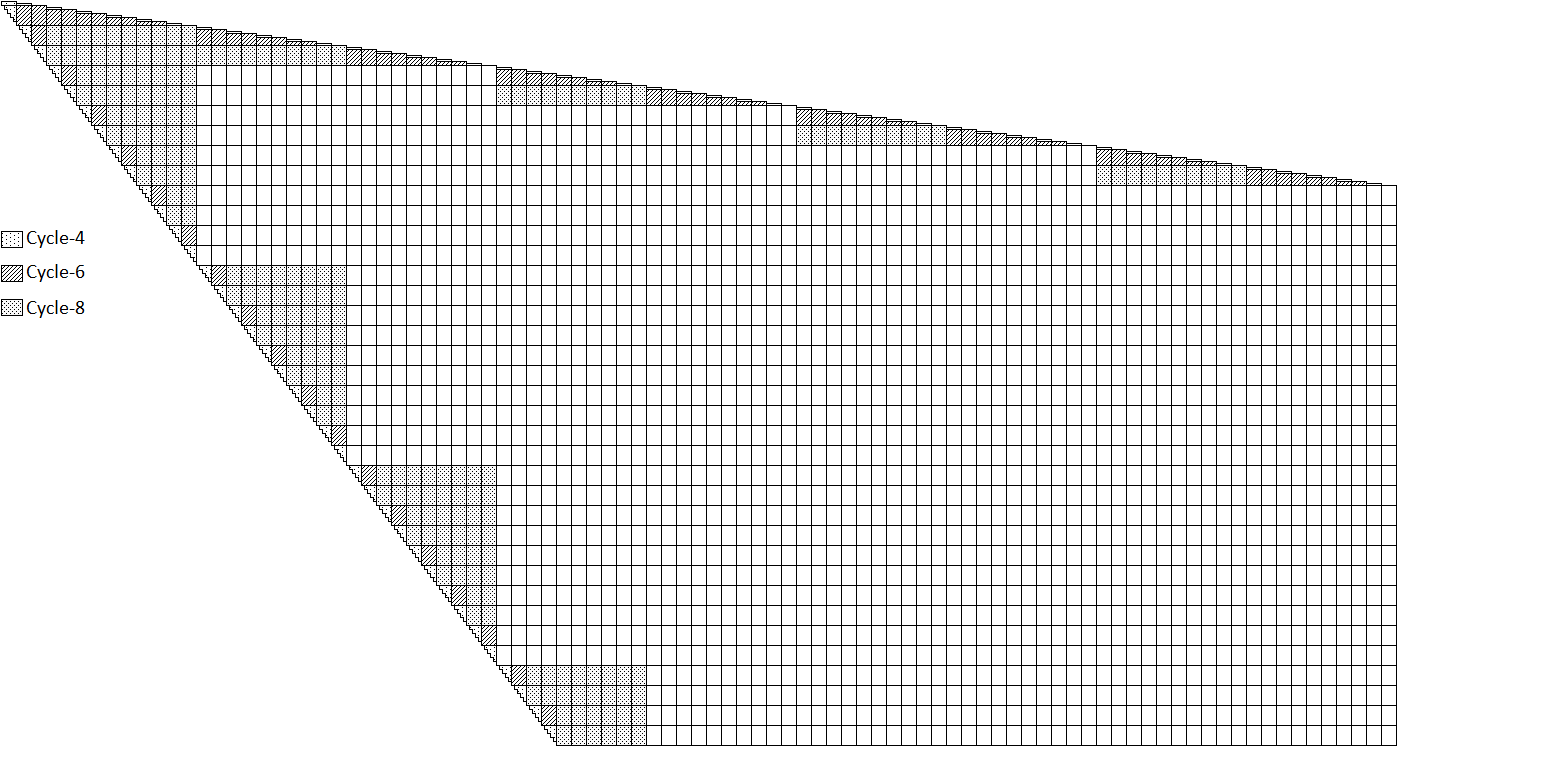}
\end{center} 
\caption{Cycle--4, Cycle--6, and Cycle--8 regions with $J = 3$ and $K = 6$}
\label{fig:Cycle468}
\end{figure}

Similar horizontal and vertical repeating patterns can be seen for Cycle--6 and Cycle--8 regions in Figure \ref{fig:Cycle468}. Making use of these patterns, one can express  $\rho(i,j)$ of an entry $(i,j)$ as a function. We introduce valid inequalities for MDD based on the cycle region information of the entries in the reduced rectangle $R$.


\begin{proposition}\label{prop3}
Let $(i,j) \in R$, i.e., $i \in \{m-r_{cr}, ..., m\}$ and $j \in \{n-c_{cr}, ..., n\}$ and let $\rho(i,j)$ represent the cycle region of the entry. Let $S$ denote the number of subblocks that intersects with $R$ and let $B_{s}$, $s \in \{1, ..., S\}$ represent the set of $(i,j)$ entries in subblock $s$. 

\begin{enumerate}
\item[(1)] If $\rho(i,j) < T$, then constraint 
\vspace{-3mm}
\begin{equation}
 X_{ij} = 0 \label{valid1}
\end{equation}
is valid.
\item[(2)] If $T = 8$ and $(i,j) \in B_s$  with $\rho(i,j) =$ 8 or 10, then constraints 
\vspace{-3mm}
\begin{equation}
\sum_{i =1}^{J - 1}\sum_{((k-1)(J - 1) + i, j) \in B_s}X_{(k-1)(J - 1) + i,j} \leq 1, \ \ \ k \in \{1, ..., K-1\}  \label{valid2}
\end{equation}
are valid.

\item[(3)] If $T = 10$ and $(i,j) \in B_s$  with $\rho(i,j) = 10$, then constraint
\vspace{-3mm}
\begin{equation}
 \sum_{(i,j) \in B_s}X_{ij} \leq 1 \label{valid3}
\end{equation}
is valid.
\end{enumerate}

\end{proposition}

\emph{Proof.} Let us consider each claim separately.

\begin{enumerate}
\item[\emph{(1)}] There cannot be cycles of length smaller than the girth $T$. If $X_{ij} = 1$, then we have a cycle of length $\rho(i,j) < T$, which is not desired. Hence, $X_{ij} = 0$ in this case.

\item[\emph{(2)}] If $T = 8$, then there should not be any cycles of length 6. Let us consider a subblock with cycle region 8 or 10, which is subdivided into $(K- 1)$ equal $subpieces$ each includes $(J-1)$ rows. In Figure \ref{fig:Cycle6on8}, we give an example for Cycle--8 subblock with $J = 3$ and $K = 6$ where we have $(K - 1)= 5$ subpieces each having $(J - 1) = 2$ rows.  As seen in figure, a cycle of length 6 forms when there is more than one nonzero entry in a subpiece. 

\begin{figure}[!h]
\begin{center}
	\includegraphics[width=0.6\columnwidth]{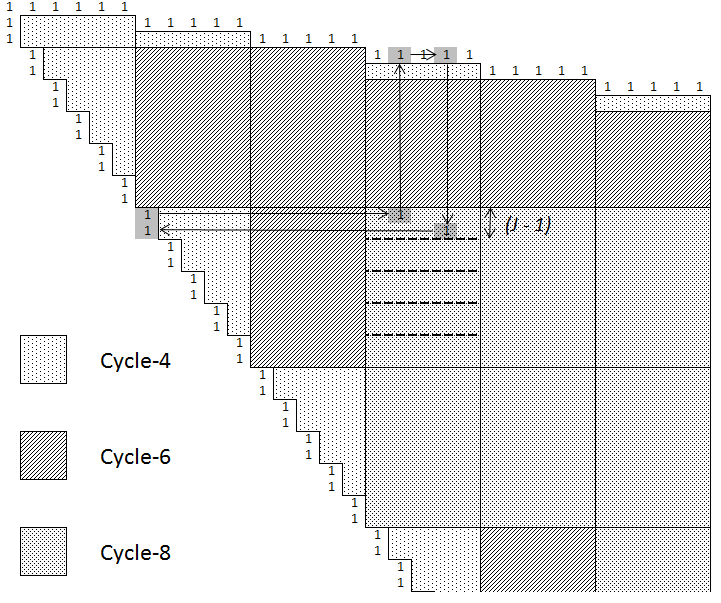}
\end{center} 
\caption{A cycle of length 6 on Cycle--8 region with $J = 3$ and $K = 6$}
\label{fig:Cycle6on8}
\end{figure}

A similar case appears for Cycle--10 subblocks. Hence, constraints (\ref{valid2}) are valid, since they force to have at most one nonzero entry in each subpiece when cycle region of the subblock is either 8 or 10.

\begin{figure}[!h]
\begin{center}
	\includegraphics[width=0.95\columnwidth]{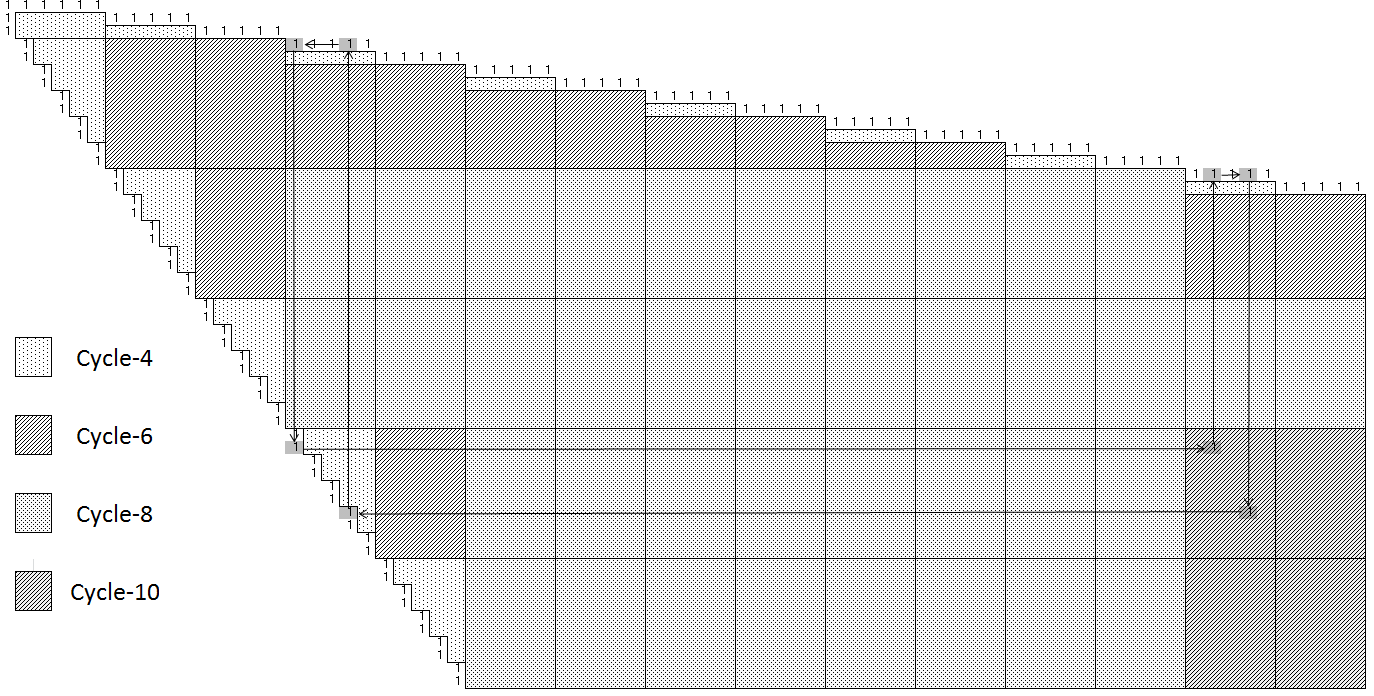}
\end{center} 
\caption{A cycle of length 8 on Cycle--10 region with $J = 3$ and $K = 6$}
\label{fig:Cycle8on10}
\end{figure}

\item[\emph{(3)}] A cycle of length 8 is not allowed when $T = 10$. However, when there is more than one nonzero entry in a subblock with cycle region 10, there is a cycle of length 8 as given in Figure \ref{fig:Cycle8on10}. Constraint (\ref{valid3}) is valid, since it bounds the number of nonzero entries from above with 1. $\square$\\

\end{enumerate}

\vspace{-5mm}
\begin{proposition}\label{prop6}
Let $z^*$ be the optimum objective value of MDD and $z_f^*$ be the optimum objective value of MDD when variables are fixed with the $extended$ mode. Let $\tau$ be defined as in Proposition \ref{prop5}. Assume there exists a  $(J, K)$--regular code with dimensions $(m, n)$, then

\begin{enumerate}
\item [(1)] $0 = z^* = z_f^*$ if $T > \tau$,
\item [(2)] $0 = z^* \leq z_f^*$ if $T \leq \tau$. 
\end{enumerate}

\end{proposition}

\emph{Proof.} For any dimensions $(m, n)$, we have $z^* \leq z_f^*$, since we fix some $X_{ij}$ variables in the $extended$ mode.  If there exists a  $(J, K)$--regular code, then there is an optimal solution with objective value $z^* = 0$. We know from Proposition \ref{prop5}  when  $T > \tau$, a  $(J, K)$--regular code can be expressed as in Figure \ref{fig:Reorder}, which coincides with the case in the $extended$ mode. Hence, we have  $ z_f^* = z^* = 0$.

In MDD if  $\rho(i,j) \geq T$, then $X_{ij}$ can be nonzero without harming the girth $T$. When $T \leq \tau$, there are $(i,j) \in S$ in Figure \ref{fig:Reorder} with $\rho(i,j) \geq T$ and they are fixed to zero, since we fix all entries in the region $S$ to zero in the $extended$ mode. Then,  we have $0 = z^* \leq z_f^*$ in this case. $\square$\\



\subsubsection{Modified Progressive Edge Growth Algorithm} \label{sec:PEG}

The last improvement to our BC algorithm is to introduce a starting solution for an initial upper bound. For this purpose, we adapt an existing algorithm from the literature known as Progressive Edge Growth (PEG) algorithm \cite{HEA01}. We modify this algorithm for our problem by starting PEG from a partial initial solution generated by our fixing algorithm given in Algorithm 4.  We also update PEG such that the  generated solution has girth at least $T$. Time complexity of Algorithm 6 is the same with the original PEG, which is $\mathcal{O}(|V||E| + |E|^2)$. In Algorithm 1, we set an upper bound by applying Algorithm 6 at step $(I.3)$. 

\vspace{-2mm}

\begin{onehalfspace}
\begin{center}
\footnotesize
$
\begin{tabular}{ll}
\textbf{Algorithm 6:} (Modified PEG) \\
\hline
\vspace{-4mm}\\
\textbf{Input:} $(m, n)$ dimensions, $\mathbf{dv}$ and $\mathbf{dc}$ vectors,  $T$ value  \\ 
\hline
\vspace{-4mm}\\
0. Initialize $\mathbf{X} \leftarrow \mathbf{0}$, $\mathbf{dv^{c}} \leftarrow \mathbf{0}$, $\mathbf{dv^{s}} \leftarrow \mathbf{dv}$ and $\mathbf{dc^{s} \leftarrow \mathbf{dc}}$, $\mathcal{I} \leftarrow \mathbf{0}$ \\  
1. Apply Algorithm 4 and update slacks \\ 
 \hspace{10pt}  $dv_j^{s} \leftarrow dv_j^{s} - \sum_{i}X_{ij}$ for all $j$ and  $dc_i^{s} \leftarrow dc_i^{s} - \sum_{j}X_{ij}$ for all $i$\\
 \hspace{10pt} and current degrees $dv_j^{c} \leftarrow \sum_{i}X_{ij}$ for all $j$ \\
2.  \textbf{For} $j \in \{1, ..., n\}$ set $\mathcal{I} \leftarrow \mathbf{0}$\\
3. \hspace{10pt} \textbf{For} $k \in \{0, ..., dv_j^{c}\}$ \\
4.  \hspace{20pt} \textbf{If} $k = 0$, \textbf{Then} set $X_{i'j} = 1$ for $i' = \argmax_i\{dc_i^{s}\}$\\
5.  \hspace{20pt} \textbf{Else} apply BFS from $v_j$ to reach check nodes, let tree has depth $l$  \\
6.  \hspace{40pt} \textbf{If} $2l \geq T$  or  $|\mathcal{N}_j^{l}| \leq m$, let $\mathcal{I}$ is incidence vector for $\mathcal{N}_j^{l}$\\
     \hspace{70pt} set $X_{i'j} = 1$ for $i' = \argmax_i\{(1 - \mathcal{I}_{c_i}) dc_i^{s}\}$ \\
7.  \hspace{20pt} \textbf{End If} \\
8.  \hspace{20pt} Update $dv_j^{c}$, $dv_j^{s}$,  $dc_i^{s}$ as in Step 1 \\ 
9.  \hspace{10pt} \textbf{End For}\\
10. \textbf{End For}\\
\hline
\vspace{-4mm}\\
\textbf{Output:} An initial solution for MDD \\
\hline
\end{tabular}
$
\end{center}
\end{onehalfspace}
\vspace{5mm}

In Algorithm 6, $\mathbf{dv}$ and $\mathbf{dc}$ are the target degree vectors for variable and check nodes, respectively. Let deviation from the target degrees for variable and check nodes be given by slack vectors  $\mathbf{dv^{s}}$ and $\mathbf{dc^{s}}$, and the current degrees of variable nodes be listed in vector $\mathbf{dv^{c}}$. Moreover, $\mathcal{N}_j^{l}$ represents the set of all check nodes that can be reached from $v_j$ with a tree of depth $l$. Hence, the set  $\mathcal{N}_j^{l} \setminus\mathcal{N}_j^{l-1}$ collects the check nodes that are reached at the $l$th step from $v_j$ for the first time. We can represent the check nodes in the set $\mathcal{N}_j^{l}$ with an incidence vector $\mathcal{I}$ as $\mathcal{I}_{c_i} = 1$ if $c_i \in \mathcal{N}_j^{l}$ and zero otherwise.


Starting from the solution provided by Algorithm 4, PEG adds an edge $(i,j)$, i.e., $X_{ij} = 1$, if this edge does not form a cycle ($|\mathcal{N}_j^{l}| \leq m$) or the length of the cycle created is greater or equal to $T$ (Step 6). For edge assignment, the algorithm picks $c_i$ having the maximum slack value $dc_i^{s}$ in order to fit the target degree $dc_i$. The generated solution is feasible for MDD, since it has girth at least $T$.

\section{Computational Results} \label{ComputationalResults}

The computations have been carried out on a computer with 2.0 GHz Intel Xeon E5--2620 processor and 46 GB of RAM working under Windows Server 2012 R2 operating system.  
In the computational experiments, we use CPLEX 12.6.2 to test the performance of BC algorithm and evaluate how different improvement strategies to BC algorithm given in Section \ref{sec:Improvements} affect the results. We implement all algorithms in the C++ programming language. We summarize the solution methods in Table \ref{tab:sosm}.

\begin{onehalfspace}
\begin{center}
\captionof{table}{Summary of solution methods}
    \label{tab:sosm}
\begin{tabular}{cccc} 
\hline
Method & Mode & Valid Inequalities & Modified PEG \\ 
\hline
BC$_0$ & -- & -- & --  \\
BC$_1$ & $basic$ & -- & -- \\
BC$_2$ & $extended$ & -- & --  \\
BC$_3$ & $extended$ & $\surd$ & -- \\
BC$_4$ & $extended$ & $\surd$ & $\surd$ \\
\hline
\end{tabular}
\end{center}
\end{onehalfspace}

In BC$_0$, we apply the BC algorithm in Algorithm 1 without improvement techniques, i.e., we exclude steps $(I.1)-(I.3)$. Algorithm 1 includes Algorithm 2 and 3 to separate integral and fractional solutions, respectively. In CPLEX, we implement Algorithm 2 using $LazyConstraintCallback$ and Algorithm 3 with $UserConstraintCallback$ routines. We utilize default branching settings of CPLEX. In BC$_1$ method, we apply step $(I.1)$  to fix the first row and column of $\mathbf{H}$ matrix in the $basic$ mode. In BC$_2$ method, step $(I.1)$  fixes $r_{cr}$ rows and $c_{cr}$ columns in the $extended$ mode (see Section \ref{sec:VariableFixing}). In BC$_3$ method, we apply step $(I.1)$  in  the $extended$ mode and step $(I.2)$  adds valid inequalities that are explained in Section \ref{sec:ValidInequalities}. Finally in BC$_4$ method, step $(I.1)$  runs in the $extended$ mode, step $(I.2)$  adds valid inequalities and step $(I.3)$  provides an initial solution with modified PEG (Algorithm 6). 

We list the parameters used in the computational experiments in Table \ref{tab:locp}. We generate $(3, 6)-$regular $\mathbf{H}$ matrices with girth values $T = 6, 8$ or 10 in our experiments. We try nine different $(m, n)$ dimensions from $n= 20$ to 1000. We report the results that CPLEX found in 3600 seconds time limit.

\begin{onehalfspace}
\begin{center}
\captionof{table}{List of computational parameters}
    \label{tab:locp}
\begin{tabular}{c l}
\hline
\multicolumn{2}{c}{\textit{Parameters}} \\
\cline{1-2}
$(J, K)$ & $(3, 6)-$regular codes \\
$(m, n)$  & (10, 20), (15, 30), (20, 40), (30, 60), \\
 & (40, 80), (100, 200), (150, 300), (250, 500), (500, 1000)   \\
$T$ & 6, 8, 10 \\
Time Limit & 3600 secs \\
\hline
\end{tabular}
\end{center}
\end{onehalfspace}

\vspace{3mm}

From Table \ref{tab:BC$_0$} to \ref{tab:BC$_3_4$}, column ``$z$" is the objective function value of MDD and column ``$z_l$" is the best known lower bound found by CPLEX within the time limit. For each of the methods, we have an initial feasible solution (an upper bound) with objective value  $z^i_u$. In BC$_0$ method, $\mathbf{H} = \mathbf{0}$ is a trivial solution providing an initial upper bound. In methods from BC$_1$ to BC$_4$ an initial feasible solution is obtained from variable fixing (see Section \ref{sec:VariableFixing}) or modified PEG heuristic (see Section \ref{sec:PEG}). Computational time in seconds is given with column ``CPU (secs)" and percentage difference among $z_l$ and $z$ is under column ``Gap (\%)". In column ``Lazy" we show number of cuts added to MDD using Algorithm 2, whereas column ``User" is the number of cuts added to MDD with Algorithm 3. 

As discussed in Section \ref{sec:MathematicalFormulations}, we have a $(J, K)$--regular code if $z_l = z = 0$. We can conclude that it is not possible to have a $(J, K)$--regular code with given $(m, n)$ and the girth $T$ when we have $z \geq z_l > 0$ (see Proposition \ref{prop8}). In Table \ref{tab:BC$_0$}, we can see that BC$_0$ can find a $(3, 6)-$regular code for 8 instances when $T = 6$. As $T$ and $n$ increase, BC$_0$ method cannot improve initial upper bound $z^i_u$. For $T = 8$ and $T = 10$, we observe that the number of lazy and user cuts added to MDD gets smaller as $n$ gets larger.  This is because adding a cut takes more time as $n$ increases, which causes the algorithm to generate fewer cuts within the given time limit.

\begin{onehalfspace}
\begin{center}
\footnotesize
\captionof{table}{Computational results for BC$_0$}
    \label{tab:BC$_0$}
\begin{tabular}{ccccccccc}
    \hline
 &   & &  & & CPU & Gap & \multicolumn{2}{c}{\# Cuts}  \\ \cline{8-9}
$T$ & $n$ & $z_l$ &  $z$ & $z^i_u$ & (secs) &  (\%) & Lazy & User  \\
    \hline
6 & 20 & 0 & 20 & 120 & $time$ & 100 & 7399 & 0 \\
& 30 & 0 & 0 & 180 & 13.80 & 0 & 5784 & 0 \\
& 40 & 0 & 0 & 240 & 0.39 & 0 & 331 & 0  \\
& 60 & 0 & 0 & 360 & 0.45 & 0 & 184 & 0  \\
& 80 & 0 & 0 & 480 & 0.41 & 0 & 94 & 0  \\
& 200 & 0 & 0 & 1200 & 1.06 & 0 & 238 & 0 \\
& 300 & 0 & 0 & 1800 & 2.62 & 0 & 165 & 0 \\
& 500 & 0 & 0 & 3000 & 4.72 & 0 & 114 & 0 \\
& 1000 & 0 & 0 & 6000 & 32.71 & 0 & 111 & 0 \\
8 & 20 & 0 & 62 & 120 & $time$ & 100 & 51759 & 19192 \\
& 30 & 0 & 86 & 180 & $time$ & 100 & 138018 & 9890 \\
& 40 & 0 & 240 & 240 & $time$ & 100 & 196066 & 4452 \\
& 60 & 0 & 360 & 360 & $time$ & 100 & 285614 & 2683 \\
& 80 & 0 & 480 & 480 & $time$ & 100 & 328598 & 2055 \\
& 200 & 0 & 1200 & 1200 & $time$ & 100 & 404838 & 736 \\
& 300 & 0 & 1800 & 1800 & $time$ & 100 & 327245 & 261 \\
& 500 & 0 & 3000 & 3000 & $time$ & 100 & 207064 & 61 \\
& 1000 & 0 & 0 & 6000 & 905.21 & 0 & 2458 & 2 \\
10 & 20 & 0 & 62 & 120 & $time$ & 100 & 171969 & 31649 \\
& 30 & 0 & 164 & 180 & $time$ & 100 & 393619 & 7676 \\
& 40 & 0 & 240 & 240 & $time$ & 100 & 410765 & 5554 \\
& 60 & 0 & 360 & 360 & $time$ & 100 & 554898 & 3740 \\
& 80 & 0 & 480 & 480 & $time$ & 100 & 496226 & 2465 \\
& 200 & 0 & 1200 & 1200 & $time$ & 100 & 67718 & 406 \\
& 300 & 0 & 1800 & 1800 & $time$ & 100 & 22282 & 88 \\
& 500 & 0 & 3000 & 3000 & $time$ & 100 & 11548 & 10 \\
& 1000 & 0 & 6000 & 6000 & $time$ & 100 & 87546 & 65 \\
\hline
\end{tabular}
\end{center}
\end{onehalfspace}

\vspace{3mm}

Table \ref{tab:BC$_1_2$} shows our computational results for BC$_1$ and BC$_2$. We have better initial upper bound ($z^i_u$) values compared to BC$_0$ when we implement variable fixing with the $basic$ mode in BC$_1$. We improve $z^i_u$ values more in BC$_2$ with the $extended$ mode, since we fix more entries compared to the $basic$ mode. We observe that $z_l = 1$ for  $T = 6$ and $n = 20$ in BC$_1$, which means it is not possible to have a $(3, 6)-$regular code for this dimension. BC$_1$ method is able to solve 9 instances out of 27 instances to optimality, i.e., Gap (\%) value is zero.


\begin{onehalfspace}
\begin{center}
\scriptsize
\captionof{table}{Computational results for BC$_1$ and BC$_2$}
    \label{tab:BC$_1_2$}
\begin{tabular}{ccccccccccccccccc}
    \hline
& & \multicolumn{7}{c}{BC$_1$} & & \multicolumn{7}{c}{BC$_2$} \\  \cline{3-9} \cline{11-17}
 &   & &  & & CPU & Gap & \multicolumn{2}{c}{\# Cuts}  & &  & & & CPU & Gap  & \multicolumn{2}{c}{\# Cuts}  \\ \cline{8-9} \cline{16-17}
$T$ & $n$ & $z_l$ &  $z$ & $z^i_u$ & (secs) & (\%) & Lazy & User & & $z_l$ &  $z$ & $z^i_u$ & (secs) &  (\%) & Lazy & User \\
    \hline
6 & 20 & 1 & 20 & 104 & $time$ & 95 & 3804 & 0 & & 12 & 20 & 62 & $time$ & 40 & 246 & 0 \\
& 30 & 0 & 0 & 164 & 23.11 & 0 & 7016 & 0 & & 0 & 0 & 92 & 0.10 & 0 & 2532 & 0 \\
& 40 & 0 & 0 & 224 & 0.39 & 0 & 420 & 0 & & 0 & 0 & 122 & 0.12 & 0 & 160 & 0 \\
& 60 & 0 & 0 & 344 & 0.37 & 0 & 124 & 0 & & 0 & 0 & 182 & 0.20 & 0 & 148 & 0  \\
& 80 & 0 & 0 & 464 & 0.56 & 0 & 125 & 0 & & 0 & 0 & 242 & 0.23 & 0 & 146 & 0 \\
& 200 & 0 & 0 & 1184 & 1.43 & 0 & 108 & 0 & & 0 & 0 & 602 & 0.48 & 0 & 109 & 0 \\
& 300 & 0 & 0 & 1784 & 2.31 & 0 & 87 & 0 & & 0 & 0 & 902 & 1.11 & 0 & 167 & 0 \\
& 500 & 0 & 0 & 2984 & 4.73 & 0 & 94 & 0 & & 0 & 0 & 1502 & 2.44 & 0 & 225 & 0 \\
& 1000 & 0 & 0 & 5984 & 49.23 & 0 & 110 & 0 & & 0 & 0 & 3002 & 21.83 & 0 & 165 & 0 \\
8 & 20 & 0 & 44 & 104 & $time$ & 100 & 19099 & 16644 & & 42 & 42 & 62 & 0.08 & 0 & 0 & 0 \\
& 30 & 0 & 74 & 164 & $time$ & 100 & 73701 & 8222 & & 64 & 64 & 92 & 0.33 & 0 & 244 & 0 \\
& 40 & 0 & 92 & 224 & $time$ & 100 & 131947 & 4385 & & 56 & 84 & 122 & $time$ & 32 & 2660 & 68 \\
& 60 & 0 & 344 & 344 & $time$ & 100 & 225388 & 1903 & & 12 & 80 & 182 & $time$ & 85 & 25418 & 0 \\
& 80 & 0 & 464 & 464 & $time$ & 100 & 240048 & 1703 & & 0 & 242 & 242 & $time$ & 100 & 61703 & 0 \\
& 200 & 0 & 1184 & 1184 & $time$ & 100 & 407426 & 895 & & 0 & 602 & 602 & $time$ & 100 & 229615 & 0 \\
& 300 & 0 & 1784 & 1784 & $time$ & 100 & 331382 & 487 & & 0 & 902 & 902 & $time$ & 100 & 292952 & 0 \\
& 500 & 0 & 2984 & 2984 & $time$ & 100 & 216118 & 124 & & 0 & 0 & 1502 & 1633.83 & 0 & 148866 & 0 \\
& 1000 & 0 & 0 & 5984 & 454.20 & 0 & 1386 & 6 & & 0 & 0 & 3002 & 449.31 & 0 & 1263 & 0 \\
10 & 20 & 0 & 58 & 104 & $time$ & 100 & 57480 & 80057 & & 54 & 54 & 62 & 0.09 & 0 & 0 & 0 \\
& 30 & 0 & 164 & 164 & $time$ & 100 & 242023 & 16891 & & 92 & 92 & 92 & 0.09 & 0 & 0 & 0 \\
& 40 & 0 & 224 & 224 & $time$ & 100 & 342790 & 8174 & & 122 & 122 & 122 & 0.11 & 0 & 0 & 0 \\
& 60 & 0 & 344 & 344 & $time$ & 100 & 290718 & 3953 & & 182 & 182 & 182 & 0.14 & 0 & 0 & 0 \\
& 80 & 0 & 464 & 464 & $time$ & 100 & 471767 & 5285 & & 236 & 236 & 242 & 142.56 & 0 & 3850 & 42 \\
& 200 & 0 & 1184 & 1184 & $time$ & 100 & 51505 & 675 & & 66 & 602 & 602 &  $time$ & 89 & 310451 & 1 \\
& 300 & 0 & 1784 & 1784 & $time$ & 100 & 20565 & 135 & & 0 & 902 & 902 &  $time$ & 100 & 461039 & 0 \\
& 500 & 0 & 2984 & 2984 & $time$ & 100 & 9568 & 60 & & 0 & 1502 & 1502 &  $time$ & 100 & 467420 & 0 \\
& 1000 & 0 & 5984 & 5984 & $time$ & 100 & 90273 & 91 & & 0 & 3002 & 3002 &  $time$ & 100 & 110798 & 0 \\
\hline
\end{tabular}
\end{center}
\end{onehalfspace}

\vspace{3mm}

In Table \ref{tab:BC$_1_2$}, we observe that we can solve 17 instances to optimality with  BC$_2$ method. BC$_2$ finds $z_l > 0$ for 11 instances indicating that there are no $(3, 6)-$regular codes for those dimensions. There are 7 instances such as $T = 10$ and $n = 80$ that we have $z_l = z > 0$. This means that for $n = 80$ dimension, the best possible code with the girth $T = 10$ includes $z /2 = 236 /2 = 118$ fewer ones than a $(3, 6)-$regular code (having $X_{ij} = 1$ improves MDD objective by 2).

Comparing Table \ref{tab:BC$_1_2$} and \ref{tab:BC$_3_4$}, we can see that $z^i_u$ values for BC$_2$ and BC$_3$ are the same, since we apply the $extended$ mode for both. On the other hand, feasible solution of Algorithm 6 (see Section \ref{sec:PEG}) provides better $z^i_u$ values in BC$_4$. Results show that $z$ values get better, the number of cuts added to MDD gets smaller and computational time improves on the average as we have tighter initial solutions.   

\newpage
\begin{onehalfspace}
\begin{center}
\scriptsize 
\captionof{table}{Computational results for BC$_3$ and BC$_4$}
    \label{tab:BC$_3_4$}
\begin{tabular}{ccccccccccccccccc}
    \hline
 & & \multicolumn{7}{c}{BC$_3$} & & \multicolumn{7}{c}{BC$_4$} \\  \cline{3-9} \cline{11-17}
 &   & &  & & CPU & Gap & \multicolumn{2}{c}{\# Cuts}  & &  & & & CPU & Gap  & \multicolumn{2}{c}{\# Cuts}  \\ \cline{8-9} \cline{16-17}
$T$ & $n$ & $z_l$ &  $z$ & $z^i_u$ & (secs) & (\%) & Lazy & User & & $z_l$ &  $z$ & $z^i_u$ & (secs) &  (\%) & Lazy & User \\
    \hline
6 & 20 & 12 & 20 & 62 & $time$ & 40 & 260 & 0 & & 13.9 & 20 & 26 & $time$ & 37 & 238 & 0 \\
& 30 & 0 & 0 & 92 & 0.15 & 0 & 1784 & 0 & & 0 & 0 & 8 & 0.22 & 0 & 2522 & 0 \\
& 40 & 0 & 0 & 122 & 0.14 & 0 & 160 & 0 & & 0 & 0 & 2 & 0.36 & 0 & 441 & 0 \\
& 60 & 0 & 0 & 182 & 0.20 & 0 & 160 & 0 & & 0 & 0 & 2 & 0.16 & 0 & 154 & 0 \\
& 80 & 0 & 0 & 242 & 0.24 & 0 & 148 & 0 & & 0 & 0 & 2 & 0.33 & 0 & 184 & 0 \\
& 200 & 0 & 0 & 602 & 0.55 & 0 & 109 & 0 & & 0 & 0 & 4 & 0.56 & 0 & 104 & 0 \\
& 300 & 0 & 0 & 902 & 1.02 & 0 & 167 & 0 & & 0 & 0 & 2 & 1.11 & 0 & 167 & 0 \\
& 500 & 0 & 0 & 1502 & 3.33 & 0 & 225 & 0 & & 0 & 0 & 2 & 3.05 & 0 & 207 & 0 \\
& 1000 & 0 & 0 & 3002 & 39.79 & 0 & 170 & 0 & & 0 & 0 & 4 & 29.84 & 0 & 174 & 0 \\
8 & 20 & 42 & 42 & 62 & 0.12 & 0 & 0 & 0 & & 42 & 42 & 62 & 0.13 & 0 & 0 & 0 \\
& 30 & 64 & 64 & 92 & 0.16 & 0 & 0 & 0 & & 64 & 64 & 86 & 0.13 & 0 & 0 & 0 \\
& 40 & 84 & 84 & 122 & 7.89 & 0 & 473 & 0 & & 84 & 84 & 86 & 2.59 & 0 & 367 & 0 \\
& 60 & 28 & 64 & 182 & $time$ & 56 & 55860 & 0 & & 28 & 60 & 66 & $time$ & 53 & 58432 & 0 \\
& 80 & 8 & 242 & 242 & $time$ & 97 & 95449 & 0 & & 8 & 38 & 38 & $time$ & 87 & 83615 & 0 \\
& 200 & 0 & 0 & 602 & 2181.18 & 0 & 154415 & 0 & & 0 & 0 & 16 & 1893.82 & 0 & 166949 & 0 \\
& 300 & 0 & 902 & 902 & $time$ & 100 & 280596 & 0 & & 0 & 10 & 10 & $time$ & 100 & 284583 & 0 \\
& 500 & 0 & 0 & 1502 & 614.80 & 0 & 33635 & 0 & & 0 & 0 & 10 & 1414.95 & 0 & 71447 & 0 \\
& 1000 & 0 & 0 & 3002 & 324.91 & 0 & 587 & 0 & & 0 & 0 & 12 & 384.75 & 0 & 866 & 0 \\
10 & 20 & 54 & 54 & 62 & 0.10 & 0 & 0 & 0 & & 54 & 54 & 62 & 0.13 & 0 & 0 & 0 \\
& 30 & 92 & 92 & 92 & 0.09 & 0 & 0 & 0 & & 92 & 92 & 92 & 0.11 & 0 & 0 & 0 \\
& 40 & 122 & 122 & 122 & 0.11 & 0 & 0 & 0 & & 122 & 122 & 122 & 0.17 & 0 & 0 & 0 \\
& 60 & 182 & 182 & 182 & 0.11 & 0 & 0 & 0 & & 182 & 182 & 182 & 0.13 & 0 & 0 & 0 \\
& 80 & 236 & 236 & 242 & 0.18 & 0 & 1 & 0 & & 236 & 236 & 236 & 0.17 & 0 & 0 & 0 \\
& 200 & 260 & 602 & 602 & $time$ & 57 & 100732 & 4 & & 260 & 314 & 314 & $time$ & 17 & 78306 & 16 \\
& 300 & 104 & 902 & 902 & $time$ & 88 & 273318 & 0 & & 104 & 274 & 274 & $time$ & 62 & 335686 & 0 \\
& 500 & 0 & 1502 & 1502 & $time$ & 100 & 170322 & 0 & & 0 & 174 & 174 & $time$ & 100 & 165584 & 0 \\
& 1000 & 0 & 3002 & 3002 & $time$ & 100 & 52500 & 0 & & 0 & 60 & 60 & $time$ & 100 & 47637 & 0 \\
\hline
\end{tabular}
\end{center}
\end{onehalfspace}

\vspace{3mm}

In Table \ref{tab:BC$_3_4$}, we can also compare the performance of our methods with the state--of--the--art heuristic PEG. In BC$_4$ method, $z^i_u$ values are the objective function values of PEG. BC$_4$ method can improve the solution provided by PEG for 17 instances among 27 instances. Similarly,  BC$_1$ outperfoms the PEG for 13 instances, BC$_2$ for 15 instances and BC$_3$ for 17 instances.

\begin{onehalfspace}
\begin{center}
\captionof{table}{Lower and upper bounds on the dimension $n$}
    \label{tab:LBUBonN}
\begin{tabular}{cccccc}
    \hline
 & \multicolumn{2}{c}{Proposition \ref{prop8}} & & \multicolumn{2}{c}{BC$_4$} \\  \cline{2-3} \cline{5-6}
$T$ & $r_{cr}$ & $n_{LB}$ &  &  $n_{LB}$ & $n_{UB}$   \\ 
    \hline
6 & 3 & 20 & & 20 & 30 \\
8 & 13 & 70 & & 80 & 200 \\
10 & 33 & 170 & & 300 & --- \\
\hline
\end{tabular}
\end{center}
\end{onehalfspace}

Among the methods from BC$_0$ to BC$_4$, we can see that BC$_4$ uses the smallest number of cuts on the average and solves more instances to optimality (19 instances out of 27 instances). Besides, BC$_4$ provides an evidence that there cannot be a $(J, K)$--regular code (when $z_l > 0$) for 13 instances within the given time limit. In Table \ref{tab:LBUBonN}, we compare the lower bounds on $n$ provided by Proposition \ref{prop8} and BC$_4$ for a (3,6)--regular code  with girth $T$. In  BC$_4$ method, the largest $n$ that we have $z_l > 0$ is a lower bound and the smallest $n$ that we obtain $z_l = z=0$ is an upper bound. BC$_4$ gives tighter lower bounds than Proposition \ref{prop8}. BC$_4$ can find the smallest dimension $n$ that one can generate a (3,6)--regular code with girth $T$ by applying binary search on $n$. Taking into account that code design problem is an offline problem, one can implement BC$_4$ method to construct a $(J, K)$--regular code providing sufficiently large time. 




\section{Conclusions} \label{Conclusions}

In this work, we investigate the LDPC code design problem and provide an MIP formulation for the girth feasibility problem. For the solution of the problem, we propose a branch--and--cut (BC) algorithm. We analyze structural properties of the problem and improve our BC algorithm by using techniques such as  variable fixing, adding valid inequalities and providing an initial solution using a heuristic. Computational experiments indicate that each of these techniques improves BC one step further. Among all, the method that combines all of these strategies, i.e., BC$_4$ method, can solve the largest number of instances to optimality and gives the smallest gap values on average in an acceptable amount of time. One important gain of the method is that it can provide an evidence whether there can be a  $(J, K)$--regular code with the given dimensions or not. 

In this study, our focus has been on $(J, K)$--regular codes. In telecommunication applications, irregular LDPC codes are also utilized. Hence, extending these techniques to irregular LDPC codes  can be a direction of future research.  Spatially--coupled (SC) LDPC codes are another code family which has become popular due to their channel capacity approaching error correction capability. Design of SC LDPC codes without small cycles will be a valuable contribution to future communication standards.  


\section*{Acknowledgments} 

This research has been supported by the Turkish Scientific and Technological Research Council with grant no 113M499.


\vspace{-5mm}

\end{document}